\documentstyle[12pt,epsf,amstex]{article}
\input epsf

\newcommand{\be}{\begin{eqnarray}}
\newcommand{\ee}{\end{eqnarray}}
\renewcommand{\vec}[1]{{\bf #1}}
\newcommand{\vecg}[1]{\mbox{\boldmath $#1$}}

\begin{document}

\begin{center}

{\Large\bf   Low-dimensional sisters of Seiberg--Witten effective theory}
\footnote{A contribution to Ian Kogan memorial volume.}

\vspace{0.8cm}

{\Large A.V. Smilga} \\

{\it SUBATECH, Universit\'e de
Nantes,  4 rue Alfred Kastler, BP 20722, Nantes  44307, France. }
\footnote{On leave of absence from ITEP, Moscow, Russia.}\\

\end{center}

\bigskip

\begin{abstract}
We consider the theories obtained by dimensional reduction to $D=1,2,3$ of
$4D$ supersymmetric Yang--Mills theories and calculate there the effective 
low-energy lagrangia describing moduli space dynamics --- the low-dimensional analogs
of the Seiberg--Witten effective lagrangian. The effective theories thus obtained
are rather beautiful and interesting from mathematical viewpoint. In addition, their
study allows one to understand better some essential features of $4D$ supersymmetric
theories, in particular --- the nonrenormalisation theorems. 
\end{abstract}

\newpage

\tableofcontents

\section{Introduction}
 Ian's scientific style had two attractive  features: {\it (i)} his works 
were using more often than not rather nontrivial modern mathematical constructions; 
{\it (ii)} they were always based on a solid and clear physical idea.
This text also represents an exercise (a  review of exercises) on ``physical mathematics''
 involving an interplay between purely mathematical geometric constructions and a simple
physical notion of effective lagrangian.

  Effective lagrangia/hamiltonia arise naturally in  theories involving two energy scales. 
Integrating out the ``fast'' variables (the degrees of freedom with large characteristic
excitation energy), one obtains the effective lagrangian involving only ``slow'' variables
and describing low--energy dynamics. The classical example is the Born--Oppenheimer 
effective hamiltonian
describing nuclei dynamics in a molecule and obtained after integrating out the electron
degrees of freedom. The Euler--Heisenberg effective lagrangian describing nonlinear soft photon
interactions, the effective chiral lagrangian in QCD, the Wilsonean renormalized effective lagrangian
(where modes with high frequency up to $\Lambda_{UV}$ are integrated out) all belong to this
class.

 The same concerns the famous Seiberg--Witten effective lagrangian \cite{SW}. 
Let us remind its salient features. Consider pure $4D$ ${\cal N} = 2$ supersymmetric Yang--Mills
theory. The lagrangian
written in terms of ${\cal N} = 1$ superfields 
is\,\footnote{\,Our convention is close to that of Ref. \cite{WB},
$\theta^2 = \theta^\alpha \theta_\alpha\, \bar\theta^2 = \theta_{\dot{\alpha}} \theta^{\dot{\alpha}} 
$ ,\  $\int d^2 \theta \, 
\theta^2 = \int d^2 \bar\theta \, 
\bar\theta^2 = 1$. In the following we will also use $(\sigma^\mu)_{\alpha \dot\beta} = 
\{1, \vecg{\tau}\}_{\alpha \dot\beta}$,
$(\bar\sigma^\mu)^{\dot\beta \alpha} = 
\{1, -\vecg{\tau}\}^{\dot\beta \alpha}$. 
But our Minkowski  metric
$\eta_{\mu\nu} = {\rm diag} (1,-1,-1,-1)$ differs in sign from Wess and Bagger's conventions  
 and we include the extra
factor $2$ in the definition of $V$. }
 \be
\label{N2SYM1}
 {\cal L} \ =\ \frac 1{g^2} {\rm Tr} \left \{ \int d^2 \theta \, W^\alpha W_\alpha  + 2 \int d^2\theta 
 d^2\bar\theta\,\Phi  e^{-V} \bar \Phi e^V   \right\}
 \ee
In bosonic sector, it includes the gauge field $A_\mu$ and a complex scalar $\phi$ belonging
to the adjoint representation of the gauge group,
 \be  
 \label{N2SYMcomp}
 g^2{\cal L} \ =\ - \frac 12 {\rm Tr}\{ F_{\mu\nu}^2 \} + 2{\rm Tr} \{{\cal D}_\mu 
\bar \phi {\cal D}_\mu  \phi \}
-   {\rm Tr} \{ [\bar \phi, \phi]^2 \} \ + {\rm fermions}
 \ee
The lagrangian is most economically expressed as (see e.g. \cite{Lykken})
   \be
\label{N2SYM2}
{\cal L} \ =\  \frac 1 {g^2}  {\rm Tr} \int d^2 {\theta}  d^2\tilde{\theta} \, {\cal W}^2 \ ,
 \ee
 where ${\cal W}(x_L, \theta_\alpha, \tilde{\theta}_\alpha)$ is a
 ${\cal N} = 2$ chiral superfield 
  \be
 \label{Wcal}
{\cal W} \ =\ \Phi + i \sqrt{2} \tilde{\theta}^{\alpha} W_\alpha - \frac {\tilde{\theta}^2}4 
\bar D^2 \left( e^{-V} \bar \Phi e^V \right)\ . 
  \ee
Superfields $V, W_\alpha, \Phi, \bar\Phi $ live in ordinary superspace $(x, \theta, \bar\theta)$.  
Besides the chirality conditions, $\bar D_{\dot{\alpha}}^a \, {\cal W} = 0$, the superfield (\ref{Wcal})
  satisfies the constraints
 \be
 \label{svjazW}
D^{a\alpha} D^b_\alpha \, {\cal W} \ =\  \bar D^a_{\dot{\alpha}} \bar D^{b\dot{\alpha}} \, \bar {\cal W} 
\ ,
 \ee
where $a,b=$ ({\sl nothing, tilde}) are the global $SU(2)$ indices. The superfield ${\cal W}$ can be 
naturally expressed
in the framework of 
 harmonic superspace approach (see the monography \cite{kniga} and also recent \cite{sympl} ), 
but do not themselves depend on harmonics in the chosen basis.

 This theory has (infinitely) many different classical vacua. Supersymmetric vacuum has zero energy. At the
classical level, it has zero potential energy.
Note now that the potential commutator term in (\ref{N2SYMcomp}) 
vanishes when $ [\bar\phi, \phi] = 0$, which implies that
$\phi$ belongs to the Cartan subalgebra of the corresponding Lie algebra. Factorizing over gauge
transformations, this gives $r$ physical complex  parameters ($r$ is the rank of the group) characterizing
the classical vacuum moduli space. When quantum corrections are taken into account, one could in principle 
expect the appearance of a nontrivial effective potential on the moduli space so that the energy would 
generically 
be shifted from zero. It is  specific for supersymmetric theories that quantum corrections {\it vanish}
in any order of perturbation theory. It is  specific for ${\cal N} =2$ theory that also nonperturbative
corrections to the effective potential vanish. However, the corrections to the 
kinetic part of the lagrangian need not vanish
and they do not. The relevant slow variables are $r$ complex parameters $\phi^A$ mentioned above and their 
${\cal N} =2$ superpartners involving fermions and also $r$ Abelian gauge fields. In the simplest
$SU(2)$ case, they can be combined in one ${\cal N} =2$ superfield ${\cal W} = \phi \, + \ldots\; .$\footnote{\,No spinor or matrix indices there!} The effective Seiberg--Witten lagrangian has the form
\be
\label{LSW}
  {\cal L} =  \int  d^4\theta  {F}({\cal W}) + \  {\rm c.c.}
 \ .
  \ee  
Being expressed in components, this gives a nontrivial metric on the moduli space. 
Now, $F({\cal W})$ is a nontrivial elliptic function taking account of the instanton contributions, etc.
Its large ${\cal W}$ asymptotics  is simple, however: $F({\cal W}) = \frac {{\cal W}^2}{4\pi^2} 
 \ln {\cal W}$. This takes into
account only  the perturbative corrections, which appear only at the one--loop level.

This paper is devoted  to evaluation of the effective lagrangians in the theories obtained by the 
dimensional
reduction of (\ref{N2SYMcomp}) and also by the dimensional reduction of ${\cal N}=1$ SYM theories.
 
 Let us start with discussing the latter. In four dimensions, pure SYM theories do not involve  
vacuum moduli space. The number of quantum vacua is finite there coinciding with the dual Coxeter
 number (or, 
which is the same, the adjoint Casimir operator $c_V$) of
the gauge group \cite{Coxeter}. However, moduli space appears after dimensional reduction.
Consider first the theory reduced to $(0+1)$ dimensions. In such a theory, new gauge invariants
made of the spatial components of  gauge potential appear. The simplest such invariant is
${\rm Tr} \{ A_i^2 \}$. Indeed, gauge transformation of $A_i^a$ is reduced now to multiplication
by a group matrix $O_{ab}$ and does not involve the derivative term.  
 The tree potential 
term $\propto {\rm Tr} \, [A_i, A_j]^2 $
vanishes when $ [A_i, A_j] = 0$, i.e. when $A_i$ belongs to the Cartan subalgebra. For $SU(2)$, this means
that $A_i$ can be gauge rotated to the form $c_i t^3$. Three variables $c_i$ characterize the vacuum moduli
space.  For an arbitrary gauge group, the moduli space is characterized by $3r$ parameters.
 
  Consider now the reduction to $(1+1)$ dimensions. Only two components of $A_i$ do not involve the derivative
term in their gauge transformation law and we have $2r$ physical moduli space parameter. When reducing to
$D=3$, only one component of the vector potential for each unit of rank is left, but there are also $r$  
Abelian gauge fields which are dual in three dimensions to  scalars, $\epsilon_{ijk}F_{jk} \leftrightarrow 
\partial_i  \Psi$. Thus, in three dimensions we have  $r + r = 2r$ parameters in the vacuum moduli
space.
 
 For ${\cal N} =2$ theories, the counting is basically the same, only we have to add $2r$ parameters
associated with the scalar fields. In other words, the corresponding effective lagrangia involve
 $5r$ bosonic degrees of freedom in the $1D$ case and $4r$ degrees of freedom in the $2D$ and $3D$ cases.

The paper is organized as follows. In the next section, we describe the supersymmetric quantum mechanical
models representing effective lagrangia for the theories obtained after reduction to $(0+1)$ dimensions.
They represent nonstandard (so called {\it symplectic}) supersymmetric sigma models. They are characterized
by a mismatch between the number of bosonic and fermionic degrees of freedom: for example, in the symplectic
$\sigma$ models of the first kind (obtained from ${\cal N} =1$ theories), we have 3 bosonic and 2 fermionic 
degrees of freedom for each unit of rank, while for  symplectic
$\sigma$ models of the second kind (obtained from ${\cal N} =2$ theories), we have $5r$ bosonic and $4r$
 fermionic 
degrees of freedom. We hasten to mention right now that the number of bosonic and fermionic 
{\it quantum states}
is still equal as is dictated by supersymmetry. We will  explain later why the existence of such an unusual 
${\cal N} =2$ sigma model\,\footnote{\,Our counting of ${\cal N}$ always refers to a number of minimal supercharge representations 
in a given
dimension. Thus, for $D=1$, ${\cal N}$ counts the number of {\it complex} supercharges, for $D=4$ it 
counts the number
of Weyl spinors, etc.}
(it is not K\"ahler !) does not contradict the no--go  theorem proven in 
\cite{Freedman}.   

In Sect.\ 3, we discuss 2-dimensional effective theories. The theories obtained from
 ${\cal N} =1$ $4D$ SYM  represent
conventional K\"ahler sigma models. For extended SYM, the effective theories are more interesting
 --- they enjoy ${\cal N} = 4$ supersymmetry, but are not hyper-K\"ahler
belonging to the class of so called twisted sigma models \cite{GHR}.  

Sect.\ 4 is devoted to $3D$ effective theories. They are hyper--K\"ahler sigma models. In the simplest
$SU(2)$ case, the corresponding target space is the known Atiyah--Hitchin manifold [the $(0+1)$ version
of this sigma
model describes also the dynamic of two BPS monopoles]. In the $SU(N)$ case, the target space represents
a generalized Atiyah--Hitchin manifold associated with the dynamics of $N$ BPS monopoles.\,\footnote{To avoid confusion,
we note that they are the standard monopoles of $O(3)$ Georgi--Glashow model characterized by spatial position and a single
$U(1)$ phase.}
 For an arbitrary 
gauge group, the corresponding hyper--K\"ahler manifolds (not studied by mathematicians before) 
are obtained after certain factorizations
(hyper--K\"ahler reductions) of  generalized AH manifolds.  

In Sect.\ 5 we discuss the relationship between effective lagrangia in different dimensions and
discuss in details the {\it nonrenormalization theorems} for $D=1,2,3$  and their relationship to the
conventional nonrenormalization theorems in four dimensions.

\section{\boldmath{$D=1$}\,: symplectic sigma models}

\subsection{${\cal N} =1$}

Consider the simplest example, the massless ${\cal N} =1$ $4D$ SQED 
with the lagrangian
   \be
\label{LSQED}
{\cal L} \ =\frac{1}{2e^{2}}  \int {\rm d}^2\theta\, W^2 \ +\  
   \int {\rm d}^4\theta \left[ \bar S \,e^{V} S+ 
\bar T \, e^{-V} T \right] \ ,
 \ee
($S,T$ are chiral multiplets carrying the opposite electric charges)
reduced to $(0+1)$ dimensions. 
The effective lagrangian  (determined in \cite{SQED})
depends on the gauge potentials $A_i(t)$ and their superpartners, the photino fields $\psi_\alpha(t)$,
$\alpha = 1,2$. The charged scalar and spinor fields represent fast variables that should be 
integrated over. 
Now, $A_i$, the auxiliary field $D$ and the spinor fields $\psi_\alpha$ can be combined  in a single
${\cal N} = 2$ $1D$ superfield
 \cite{my} (see also \cite{bp})\,\footnote{
We follow the notations of Ref.\cite{sympl}, $$\bar \theta^\alpha = (\theta_\alpha)^\dagger,\ \ 
\bar \theta \theta = 
\bar \theta^\alpha \theta_\alpha, \  \bar \theta \sigma_k \theta = 
\bar \theta^\alpha (\sigma_k)_\alpha^{\ \beta} \theta_\beta$$
and the indices are raised and lowered with the help of
 the invariant tensors $\epsilon^{\alpha\beta} = - \epsilon_{\alpha\beta}$.}
  \be
\label{Vcomp}
\Gamma_k &=& A_k +  \bar \theta \sigma_k  \psi  + 
\bar \psi  \sigma_k \theta  
  + \epsilon_{kjp} \dot{A}_j \bar \theta
\sigma_p  \theta + D \bar \theta \sigma_k  \theta \nonumber \\
&& +\, i(\bar\theta \sigma_k \dot{ \psi} -
\dot{\bar \psi}
\sigma_k  \theta ) \bar\theta \theta + \frac {\ddot{A}_k}4
\theta^2 \bar\theta^2 \ .
 \ee  
The field (\ref{Vcomp}) satisfies the constraints
   \be
\label{albetgam}
D_{(\alpha} \Gamma_{\beta\gamma)} \ =\ 0\,, \quad \bar D_{(\alpha}\Gamma_{\beta\gamma)} = 0\ ,
  \ee
where $\Gamma_{\alpha\beta} = \Gamma_{\beta\alpha} \ =\ 
i(\sigma_k)_\alpha^{\ \gamma} \epsilon_{\beta\gamma}  \Gamma_k$
and 
\be
D_\alpha = \frac{\partial}{\partial \theta^\alpha} + i\bar\theta_\alpha
\frac{\partial}{\partial t}\,, \;\;
\bar D_\alpha = \frac{\partial}{\partial \bar\theta^\alpha} - i\theta_\alpha
 \frac{\partial}{\partial t}
\ee
are the covariant derivatives. Actually, $\Gamma_k$ are nothing but the spatial components of the
 former $4D$ superconnections
  \be
 \label{Gammu}
\Gamma_\mu =  \frac 14 (\bar \sigma_\mu)^{\dot \beta \alpha}
\bar D_{\dot \beta}  D_\alpha  V = \ A_\mu +
\ldots \ ,
  \ee
 the covariant
background derivatives having the form $\nabla_\mu  = \partial_\mu - i\Gamma_\mu$ \cite{Siegel}.  
In one-dimensional theory, $\Gamma_k$ is gauge invariant,
$$  \delta \Gamma_{\alpha\beta} \ \sim\ (D_\alpha \bar D_\beta + D_\beta \bar D_\alpha) 
(\Lambda - \bar \Lambda) 
= 0
$$ 
as follows from (anti)chirality of $\Lambda(\bar\Lambda)$ and the $1D$ relationship 
$\{D_\alpha, \bar D_\beta \}
\ =\ 2i\epsilon_{\alpha\beta} \partial_t$.

The effective supersymmetric and gauge--invariant action  is presented in the form 
  \be
\label{act1}
S \ =\ \int dt \int d^2 \theta d^2 \bar\theta\  F(\Gamma_k)
 \ee
By construction it enjoys ${\cal N} = 2$ supersymmetry. The lagrangian is expressed in components as
follows
   \be
\label{Lcomp}
{\cal L} = \frac h {2} \dot A_k  \dot A_k + \frac {ih}{2}
\left({\bar \psi} \dot{\psi} - \dot{{\bar \psi}}  \psi 
\right) + \frac {\partial_k h}2 \epsilon_{kjp}  \dot A_j 
{\bar \psi} \sigma_p  \psi  \nonumber \\
+ \frac {hD^2}{2} - \frac {D \partial_k h}2 {\bar \psi} 
\sigma_k  \psi - \frac {\partial^2  h}8 
\bar{\psi}^2 \psi^2\ ,
 \ee
where 
 \be
\label{hF}
 h(\vec{A}) = - \frac 12 \,\partial^2 F(\vec{A})\ .
   \ee
This is a supersymmetric sigma model with conformally flat $3D$ target space,
$ds^2 = h d\vec{A}^2$. However, it is not the conventional supersymmetric sigma
model associated with the de Rahm complex. The latter has only one pair of complex
supercharges $(Q, Q^\dagger) \equiv (d, d^\dagger)$. When the target space represents
a K\"ahler manifold, one can define an extra pair of supercharges (three such extra pairs 
for hyper--K\"ahler manifolds), but in our case the target space is 3--dimensional and definitely
not K\"ahler. 

 One can also notice that the number of bosonic and fermionic degrees of freedom are not matched
in a usual way: in a conventional sigma model one has a complex fermion for each boson while
the lagrangian (\ref{Lcomp}) involves three bosonic dynamic variables and only two fermionic.
In  field theory, where each field is associated with the asymptotic quantum state, such a mismatch
would not be allowed by supersymmetry. But in supersymmetric quantum mechanics there are no problems
with the mismatch of this kind: for each nonzero eigenvalue of the hamiltonian we still have 
two bosonic and two fermionic degenerate states
   $ |n\rangle  = \Phi_n(\vec{A}, \psi_\alpha)$.\footnote{\,As is well 
known, vacuum states with zero energy need not be paired. 
In Ref. \cite{weak}, we considered SQM models with nonstandard ``weak'' supersymmetric
algebra. For such models, the exact pairing is absent also for the first excited state. But the algebra
of all models
considered in this paper is standard.}

 A reader might be somewhat confused at this point. The widely known theorem \cite{Freedman}
seems to assert that ${\cal N} = 2$ sigma models can be defined {\it only} on K\"ahler
manifolds (and  ${\cal N} = 4$ models --- only on hyper--K\"ahler manifolds). However, this theorem
was proven  under two assumptions : {\it (i)} the theory considered should be a real field theory
with  at least 2 spacetime dimensions  and {\it (ii)} the kinetic term should have a standard form
$\propto g_{ab} \partial_\mu \phi^a \partial_\mu \phi^b$. For quantum mechanics the first condition is not
satisfied and there are no restrictions whatsoever. 

 In a standard sigma model, fermions are vectors in the tangent space. In our case, they belong to the spinor
representation of $SO(3) \equiv Sp(2)$. We will call this model symplectic sigma model of the first kind
(the second kind is coming soon).   

  In our case, the function $F(\Gamma_k)$ has a certain particular form. At the tree level,
$F(\vecg{\Gamma}) = -\vecg{\Gamma}^2/(3e^2)$  and $ h=1/e^2$. This gives the lagrangian of dimensionally 
reduced photodynamics. 
Let us evaluate one--loop correction to the metric. To this end, we need to calculate the loops
of charged superfields $S,T$ in gauge background. It is convenient to do it in components. 
We choose the background $A_i = C_i +E_i t,\ \psi_\alpha = 0$ and calculate the charged scalar and fermion
loops. The corresponding contributions to the effective action have the form 
   \be
\label{detratio}
\Delta S_{\rm eff} =
-i \ln   \frac{\det^{\frac 12} (- D^2 I  + \frac i2
\sigma_{\mu\nu}  F_{\mu\nu}) }{\det^{\frac 12}\left(-D^2 I\right)} \ ,
   \ee
where $I$ is the $4 \times 4 $ unity matrix and the identity
 \be
\label{Dslash}
\left({\rm i}\,/\!\!\!\!{D}\right)^2 = - D^2 I  +
\frac{\rm i}{2} \sigma_{\mu\nu} F_{\mu\nu}\ ,
   \ee
$\sigma_{\mu\nu} = \frac 12 [\gamma_\mu, \gamma_\nu]$\ , was used. We observe that a nonzero correction is
 due solely to magnetic interactions $\propto \sigma_{\mu\nu}  \, F_{\mu\nu}$. Were the latter absent, the 
fermion
and scalar contributions would exactly cancel. 
This feature is common for all
supersymmetric gauge theories, non-Abelian and Abelian (see \cite{akhmedov} for more details). 
And this fact is related to another known fact
that, when supersymmetric $\beta$
function is calculated in the {\it instanton}
background, only the contribution  of the zero modes survives \cite{shif}.

\begin{figure}
   \begin{center}
        \epsfxsize=170pt
        \parbox{\epsfxsize}{\epsffile{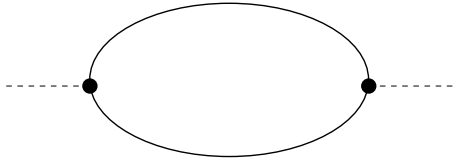}}
        \vspace{5mm}
    \end{center}
\caption{One-loop renormalization of the kinetic term in SQED. The internal lines
are Green's functions of the operator $(-D^2)$ with constant $A_i = C_i$. The vertices involve the magnetic
interaction $\propto \sigma_{0i}  \, E_i$.}
\label{loopSQED}
\end{figure}

  In the lowest order in $F_{\mu\nu}$ (or $E_i$), the contribution (\ref{detratio}) can be represented by 
the graph 
in Fig.1. Constant background $\vec{C}$ gives the ``mass'' to the charged fields and the 
Euclidean propagator
has the form $1/(\omega^2 + \vec{C}^2)$.  The calculation gives
 \be
 \label{feract}
\Delta S_{\rm eff} \ =\  -i\cdot i_{\rm Wick} \cdot i^2 \cdot  \frac 12 \cdot \left( - \frac12 \right) \nonumber \\  
\cdot E_j E_k \, {\rm Tr} \{ \sigma_{0j}\sigma_{0k} \}
\int_{-\infty}^\infty \frac{d\omega}{2\pi} \frac 1{(\omega^2 +
\vec{C}^2)^2}
\ = \  \frac {{\vec{E}}^2}{4|\vec{C}|^3}\ ,  
 \ee
where the  factor $1/2$ is the power of the determinant and the factor $-1/2$ comes from the 
expansion 
$$\ln \det \| I + \alpha \| = \ln \left[ 1 - \frac 12  {\rm Tr}\, \alpha^2 + 
\frac 13 {\rm Tr}\, \alpha^3 + \ldots \right]  \approx   - \frac 12  {\rm Tr}\, \alpha^2 $$
(Tr $\alpha = 0$ in our case).
 This immediately gives 
  \be
\label{hab}  
e^2h(\vec{C})  \ =\  1 + \frac {e^2} {2|\vec{C}|^3} + \ldots 
 \ee  

Let us discuss now non--Abelian theories. In the simplest case of the group $SU(2)$, the moduli space
involves the variables $c_k = A^3_k$ and their superpartners, which are combined in the
 superfield $\Gamma^3_k$.  The effective action has, again, the form (\ref{act1}), but the function
$F(\vecg{\Gamma}^3)$ is now different. Like in Abelian case it can be determined by calculating the loops of 
gauge and fermion fields in Abelian background $A^{\rm cl}(t) = (C_i + E_i t)t^3$ (where $C$ stands
not only for ``constant'', but also for ``Cartan''). 

The graphs are conveniently calculated in the background gauge. We represent $A_\mu = A_\mu^{\rm cl} +
{\cal A}_\mu$, where ${\cal A}_\mu$ is the quantum fluctuation and   
  and add to the Lagrangian the gauge-fixing term
\be
\label{gaugfix}
- \frac 1{2g^2} (D_\mu^{\rm cl} {\cal A}_\mu)^2\ ,
\ee
where $D_\mu^{\rm cl} = \partial_\mu - {\rm i}
 \left[A_\mu^{\rm cl}, \, \cdot \right]$.  The coefficient chosen in Eq.\,(\ref{gaugfix}) defines
the ``Feynman background gauge'', which is simpler and more
convenient than others.
Adding (\ref{gaugfix}) to the lagrangian and
integrating by parts, we obtain for
 the gauge--field--dependent part of the Lagrangian
\begin{equation}
{\cal L}_{\cal A} =
-\frac{1}{2 g^2}\, {\rm Tr} \, \left(F_{\mu\nu}^2 \right) +
 \frac{1}{g^2} \, {\rm Tr}\,
\left\{ {\cal A}_\mu \left( D^2 g_{\mu\nu}{\cal A}_\nu  - 2i
\left[F_{\mu\nu}, {\cal A}_\nu  \right] \right)
 \right\} + \ldots \ ,
\end{equation}
where the dots stand for the terms of higher order in
${\cal A}_\mu$.
The ghost part of the Lagrangian is
\begin{equation}
{\cal L}_{ghost} = - 2\, {\rm Tr} \,\left( \bar{c} D^2 c \right)
\ + {\rm higher\ order\ terms}
\end{equation}

Now we can integrate   over the quantum fields ${\cal A}_\mu$,
$c$, and  over the fermions using the relation
(\ref{Dslash}).   
We obtain  
   \be
\label{detrationab}
\delta S_{\rm eff} =
-i \ln  \left(\frac{\det^{\frac14}\left(- D^2 \, I + \frac i2
\sigma_{\mu\nu}  \left[F_{\mu\nu}, \, \cdot\right]\, \right)
\det^{1/4}\left(- D^2 I \right)}{\det^{\frac12}\left(-D^2\, g_{\mu\nu} +
2i\, \left[F_{\mu\nu}, \, \cdot\right]\right)}\right)\ .
   \ee
Again, the result would be zero in the absence of magnetic interactions. 
In this case, besides the fermion loop also the gauge field loop should be
taken into account. Nonzero commutators $\left[ F_{\mu\nu}, \, {\cal A}_\nu \right]$,
$ \left[F_{\mu\nu}, \, \lambda_\alpha \right]$
imply that the quantum fields are charged with respect to the background, i.e.
their color indices $a$ acquire the values $1,2$.

The fermion loop gives the same contribution to $S_{\rm eff}$ as in Abelian theory:
the power of the determinant is now $1/4$ rather than $1/2$ (the theory involves a Weyl
rather than Dirac fermion), but this is compensated by the extra color factor 2. One 
can be convinced that the gauge boson loop contribution involves the factor -4 compared
to the fermion one (the factor $-2$ coming from the power of determinant $-1/2$ vs. $1/4$
is explicitly seen in (\ref{detrationab})
and another factor $2$ comes from spin). This gives
 \be
\label{hnab}
g^2 h^{SU(2)}(\vec{C}) \ =\ 1 - \frac {3g^2}{2\vec{C}^2} + \ldots
 \ee
One can notice at this point that exactly the { same} graphs determine one--loop
renormalization of the effective charge in the corresponding $4D$ theories. The only difference
is that in four dimensions we have to substitute 
$$
     \int \frac{d\omega}{2\pi} \frac 1 {(\omega^2 +
\vec{C}^2)^2} \ \longrightarrow \ 
\int \frac{d^2p}{(2\pi)^4} \frac 1{(p^2 +
\vec{C}^2)^2}  \ \propto \ln \frac \Lambda {|\vec{C}|}\ .
  $$
In other words, the coefficients in (\ref{hab}), (\ref{hnab})  are rigidly related to the 
  one--loop $\beta$ function coefficients in the parent $4D$ theories. Indeed, the $\beta$ function
in non--Abelian SYM theory with $SU(2)$ gauge group involves the factor $-3$ compared to SQED.
We will return to the discussion of this point in the last section.

The metrics (\ref{hab}), (\ref{hnab}) and the relation (\ref{hF}) allow one to restore the 
corresponding prepotentials: 
  \be
e^2 F^{\rm SQED}(\vecg{\Gamma})  &=& - \frac {\vecg{\Gamma}^2}{3} + 
\frac {e^2 \ln |\vecg{\Gamma}|}{|\vecg{\Gamma}|} + ...  \nonumber \\
g^2 F^{SU(2)}(\vecg{\Gamma})  &=& - \frac {\vecg{\Gamma}^2}{3} -  
\frac {3g^2\ln |\vecg{\Gamma}|}{|\vecg{\Gamma}|} + ...
 \ee
(we posed $\vecg{\Gamma}^3 \to \vecg{\Gamma}$ in the non--Abelian case).
Consider now an arbitrary simple compact Lie group. The classical potential energy vanishes
 when $[A_j, A_k] = 0$,
which implies  that $A_j$ lies in the Cartan subalgebra (is effectively Abelian). 
This gives $3r$ bosonic variables in the effective lagrangian.
They are supplemented by $2r$ Abelian gluino variables. These variables are organized in $r$ 
superfields $\vecg{\Gamma}^{A = 1,\ldots, r}$ defined like in Eqs.\,(\ref{Vcomp}), (\ref{albetgam}). 
$\vecg{\Gamma}^A$ represent dimensionally reduced  Abelian superconnections. Thus, the effective
lagrangian has the form\,\footnote{\,It is interesting that 
such a lagrangian describes also the dynamics of $r$ extremal 
Reissner--Nordstr\"om
black holes (representing classical solutions in ${\cal N} =2$ $4D$ supergravity) \cite{Strom}.} $\int d^4\theta F(\vecg{\Gamma}^A)$ 
and the only question is what is the function
$F(\vecg{\Gamma}^A)$. Again, we have choose an Abelian gauge field background and perform the calculation
over quantum fields. The latter must have nonzero commutators with the background. They 
are classified according to the {\it roots} of the corresponding Lie algebra. Actually, we have to add the 
contributions of the loops
corresponding to each such (positive) root. The result is (see \cite{BO} for more details) 
  \be
\label{Fsumj}
g^2 F(\vecg{\Gamma}^A) \ =\   - \sum_j \left[  \frac 2{3c_V}
\left(\vecg{\Gamma}^{(j)}\right)^2 +  
\frac {3g^2}{|\vecg{\Gamma}^{(j)}|} \ln |\vecg{\Gamma}^{(j)}|\right]
\ ,
 \ee 
where $\vecg{\Gamma}^{j} = \alpha_j(\vecg{\Gamma}^A)$ and $\alpha_j$ are the roots. For example, for $SU(3)$
we have the sum of three terms with 
 \be
\label{formSU3}
\hspace{-5mm}
\alpha_1(\vecg{\Gamma}^A) = \vecg{\Gamma}^3\ , \ \ \ \  \alpha_2(\vecg{\Gamma}^A) = 
\frac{-\vecg{\Gamma}^3 + \sqrt{3} 
\vecg{\Gamma}^8}2\ , \ \ \ \  
\alpha_3(\vecg{\Gamma}^A) = \frac{\vecg{\Gamma}^3 + \sqrt{3} \vecg{\Gamma}^8}2\ . 
   \ee

\subsection{${\cal N} = 2$}
The same program can be carried out for SQM models obtained by dimensional reduction from ${\cal N} =2$ 
$4D$ theories. Consider first Abelian theory. ${\cal N} =2$ SQED has the same charged matter 
content as ${\cal N} =1$ theory, but involves an extra neutral chiral multiplet $\Phi$.
The lagrangian acquires two new terms 
 \be
\label{delLN2}
 \Delta L \ =\  \int d^4\theta \, \bar \Phi \Phi +\ 
\left[\sqrt{2}e \int d^2\theta \, \Phi ST + {\rm c.c.} \right] \ .
   \ee
The lowest component of $\Phi$ 
 gives two extra degrees of freedom in the vacuum moduli space, which becomes 
thereby 5--dimensional.\footnote{\,The moduli can be represented as spatial 
components of the gauge potential in $6D$ SQED, from 
which the
${\cal N} =2$  $4D$ theory is obtained by dimensional reduction.} The vector superfield $V$ and the
 chiral superfield
$\Phi$ can be unified in a single ${\cal N}\!=\!4$ (in SQM sense) harmonic gauge superfield and the effective
 lagrangian
can be formulated in the terms of the latter \cite{sympl}. We use here a more conventional approach
using ${\cal N} =2$ superfields. The effective action depends on $\Gamma_J = (\vecg{\Gamma}, \sqrt{2} 
\, {\rm Re} \{\Phi\}, \sqrt{2} \, {\rm Im} \{ \Phi \})$
(interpreted as superconnection in the ``grandmother'' $6D$ theory) and
must have the form
 \be
 \label{act5}
S = \int dt \int d^2 \theta d^2 \bar\theta \ {\cal K} (\vecg{\Gamma}, \bar\Phi, \Phi)\, .
 \ee
Now, ${\cal N} =2$ symmetry is manifest here. The action (\ref{act5})  is invariant under 
{\it additional} ${\cal N}=2$ supersymmetry transformations
 \be
\label{trans}
\delta \bar\Phi &=& \frac {2i}3 \epsilon^\alpha (\sigma_k)_\alpha^{\ \beta}
 D_\beta \Gamma_k\,, \nonumber \\
\delta \Phi &=&  \frac {2i}3 \bar \epsilon_\alpha (\sigma_k)_\beta^{\ \alpha}\bar D^\beta \Gamma_k\,,
\nonumber \\
\delta \Gamma_k &=& -i\epsilon^\alpha (\sigma_k)_\alpha^{\ \beta} D_\beta \Phi -
i \bar \epsilon_\alpha (\sigma_k)_\beta^{\ \alpha} \bar D^\beta \bar \Phi\,,
 \ee
provided that 
  \be
\label{harm5}
  \frac {\partial^2{\cal K}}{\partial \Gamma_k^2} +
2 \frac {\partial^2{\cal K}}{\partial \bar\Phi \partial \Phi} \equiv \frac  {\partial^2{\cal K}}
{\partial \Gamma_J^2}
 \ =\ 0\ ,
  \ee
i.e. ${\cal K}$ is a 5--dimensional harmonic function \cite{DE}. Unifying $\vec{A}$ and $\phi, \bar\phi$
in a single 5-dimensional vector $A_J$ and two spinors from the multiplets $\Gamma_k$ and $\Phi$
in a single 4--component complex spinor $\eta_\alpha$ lying in the fundamental (spinor) representation 
of $SO(5) \equiv Sp(4)$, we can write the following component expression for the lagrangian \cite{N2}
   \be
\label{LO5}
 {\cal L} \ = \ h \left[ \frac 12 \dot{A}^2_J + \frac i2 (\bar\eta
\dot{\eta} - \dot{\bar\eta} \eta ) \right] + \frac i2
\partial_J h \dot{A}_K \ \bar\eta \sigma_{JK} \eta + \nonumber \\
\frac 1{24} \left( 2 \partial_J \partial_K h - \frac 3h \partial_J h
\partial_K h \right) \left( \bar\eta \gamma_J \eta \ 
\bar\eta \gamma_K \eta - \eta C \gamma_J \eta \ 
\bar \eta  \gamma_K C \bar \eta \right)\ ,
   \ee
where $\gamma_K$ are 5--dim Dirac matrices, $\sigma_{JK} = (1/2) (\gamma_J \gamma_K - \gamma_K
\gamma_J )$ and $C$ is the antisymmetric matrix of charge conjugation, $C\gamma_J^T = -\gamma_J C$.
The metric $h$ is related to  ${\cal K}$ as $h = -(1/2) \partial^2{\cal K}/\partial \vec{A}^2$.
 
The lagrangian (\ref{LO5}) describes a sigma model defined on a conformally flat  5--dimensional target space.
We will call it symplectic sigma model of the second kind. A generalized symplectic model of the second kind
depends in this approach on  $r$ sets of  ${\cal N} = 2$ superfields 
$\Gamma_J \equiv (\vecg{\Gamma}^A, \Phi^A, \bar\Phi^A)$. The action
    \be
 \label{act5r}
S = \int dt \int d^2 \theta d^2 \bar\theta \ {\cal K} (\Gamma^A_J)\, .
    \ee
enjoys extended  ${\cal N} = 4 $ supersymmetry  provided the following generalized harmonicity 
conditions are satisfied \cite{sympl}
   \be
\label{harm5r}
\frac {\partial^2 {\cal K}}{\partial \Gamma_I^{A} \partial \Gamma_I^{B}}
\ =\ 0\ , \ \ \ \ \ \ \ \ \ \
 \frac {\partial^2 {\cal K}}{\partial \Gamma_I^{[A} \partial \Gamma_J^{B]}}
\ =\ 0\,.
    \ee

In Abelian case,  the effective action has the form (\ref{LO5}) with the {\it same} metric $h$ as
in the ${\cal N} = 1$ $4D$ SQED case discussed above. Indeed, we can choose the background with zero $\phi$
in which case the effective action is given by the  graph drawn in Fig.\ref{loopSQED}. Now, $O(5)$ invariance
dictates that the metric has the  form (\ref{hab}) also in a generic background $C_J$ with $\vec{C}^2$
being substituted by $C_J^2$. The prepotential can be chosen as 
  \be
\label{prepot}
e^2{\cal K} \ =\ - \frac {R^2}{3} + \frac {\rho^2}2 \ +\ 
\frac {e^2}{R} \ln \left(R + \sqrt{R^2 + \rho^2} \right)
\ ,
   \ee
where $R^2 = \vecg{\Gamma}^2$ and $\rho^2 = 2\bar\Phi \Phi$. Note that
${\cal K}$ need not be and is not $O(5)$ invariant.  

In non-Abelian ${\cal N} = 2$ SYM theory with $SU(2)$ gauge group, the calculations are readily done
in the same way as before. The only modification is that there are two Weyl fermions now and 
an additional adjoint scalar. The ghost determinant is exactly canceled
by the adjoint scalar determinant and we obtain
   \be
\label{detratioN2}
\delta S_{\rm eff} =
-i \ln  \left(\frac{\det^{\frac 12}\left(- D^2 \, I + \frac i2
\sigma_{\mu\nu}  \left[F_{\mu\nu}, \, \cdot\right]\, \right)}
{\det^{\frac 12}\left(-D^2\, g_{\mu\nu} +
2i\, \left[F_{\mu\nu}, \, \cdot\right]\right)}\right)\ .
   \ee
This gives the expression 
    \be
\label{hnabN2}
g^2 h^{SU(2)}_{{\cal N}=2}(C_J) \ =\ 1 - \frac {g^2}{|C_J|^3} 
 \ee
 for the metric. The respective coefficients in the correction in Abelian and non-Abelian
case conform with the respective coefficients in the corresponding $4D$ beta functions.

The structure of the expressions (\ref{hnabN2}) and    (\ref{hnab}) is similar, but there is one essential
difference. Eq.\,(\ref{hnabN2}) does not involve dots! The expression for the metric is {\it exact} and
higher loop corrections vanish. The proof of this {\it nonrenormalization theorem} is simple. Dimensional
counting tells us that an  n--loop correction to the metric
should be proportional to $(A_J A_J)^{-3n/2}$. But this is not  harmonic  for $n \geq 2$ and is excluded
by supersymmetry requirements. We will discuss the relationship of this nonrenormalization theorem to
the $4D$ nonrenormalization theorem (in ${\cal N}=2$ theories two and higher loop contributions to the 
beta function vanish) in  Sect. 5.

We want to emphasize that the absence of the corrections to the metric does not mean the absence of
the corrections to the effective lagrangian. The latter  involves  higher derivative corrections, which 
do not vanish
neither at one--loop nor at two and higher loop level \cite{sestry}.
Thus, singularity of the metric at $A_J^2 = 0$ has no great physical meaning: anyway the effective lagrangian
involves uncontrollable higher--derivative corrections  there. 

The effective lagrangian can also be found  for an arbitrary gauge group. Again, we have to sum over all
positive roots. The prepotential is
 \be
 \label{prepotr}
g^2 {\cal K}  \ =\  - \sum_j \left\{ \frac 2{3c_V} \left[ 
\left(R^{(j)}\right)^2 -
\frac 32 \left(\rho^{(j)}\right)^2 \right] + \right. \nonumber
\\ \left. 
\frac {2g^2}{R^{(j)}} \ln \left[ R^{(j)} + 
\sqrt{\left(R^{(j)}\right)^2 + \left(\rho^{(j)} \right)^2}
 \right] \right\}\ ,
   \ee
where $\left(R^{(j)}\right)^2 = \left(\vecg{\Gamma}^{(j)}\right)^2$,
 $\left(\rho^{(j)}\right)^2 = 2 \bar \Phi^{(j)}
\Phi^{(j)}$ and $\vecg{\Gamma}^{(j)} = \alpha_j(\vecg{\Gamma}^A)$, 
 $\Phi^{(j)} = \alpha_j(\Phi^A)$.

\section{\boldmath{$D=2$}\,: K\"ahler and twisted}

\subsection{${\cal N}=1:$ Unfolding  the ring}
Consider first Abelian theory. As was noted, in two dimensions we have two rather than three
moduli representing the components of the gauge potential in the reduced dimensions. 
The bosonic part of the effective lagrangian can be evaluated in the same way as in the $1D$ case
by calculating the loop diagram in Fig.\ref{loopSQED}. The only difference is that the loop integral is
two--dimensional now. We obtain 
  \be
\label{LbosD2}
e^2 {\cal L}_{\rm eff}^{\rm bos} \ =\ \frac {(\partial_\alpha A_j)^2}2  \left[ 1 + 
\frac {e^2}{2\pi A_j^2} + \ldots 
\right]\ ,
 \ee
$\alpha = 1,2$ and $j = 1,2$. 
This describes a sigma model on a 2--dimensional target space. One can, of course, introduce 
the complex 
coordinate $\sigma = (A_1 + iA_2)/\sqrt{2}$. 

The full effective lagrangian involves besides $A_j$ their supersymmetric partners --- 
 two-component photino fields. We see that in this case there is perfect matching between the 
number
of bosonic and fermionic degrees of freedom. Actually, the Alvarez--Gaume--Freedman theorem 
\cite{Freedman}
dictates that the only two--dimensional ${\cal N} = 2$ supersymmetric theory with standard 
sigma--model
kinetic term like in (\ref{LbosD2}) is the supersymmetric K\"ahler sigma model. The K\"ahler
 potential 
${\cal K}$ (${\cal L} = \int d^4\theta {\cal K}$ ) can be 
restored from the metric. In the case under consideration it can be chosen as
 \be
\label{Kab}
e^2 {\cal K}(\bar\Phi, \Phi) \ =\ \bar\Phi \Phi + \frac {e^2}{4\pi} \ln \Phi 
\ln \bar\Phi \ .
 \ee 
Now, $\Phi$ is a chiral superfield that is related to the gauge--invariant  
superconnections $\Gamma_j$ in reduced
dimensions in the  following way. Consider the superfield 
$\Sigma  = (\Gamma_1 + i\Gamma_2)/\sqrt{2}$. From the definition (\ref{Gammu}) and
the $2D$ anticommutation relations between $D_\alpha$ and $\bar D_{\dot{\alpha}}$,
we deduce that $\Sigma$ satisfies the constraints
   \be
  \label{contwist}   
 \bar {\cal D}_1 \Sigma \ =\ 
{\cal D}_2 \Sigma  = 0\ 
   \ee
and represented a so called {\it twisted} chiral multiplet. It 
  differs from the standard one by a pure convention: 
$\Sigma$ is obtained from $\Phi$ by interchanging
$\theta_2$ and $\bar\theta^2$. This means that the change $\Phi \to \Sigma$ in any
 standard action 
involving $\Phi$ would not change anything except the sign due to the change of sign 
of $d^4\theta$. For example, the tree Lagrangian is expressed as
  \be
\label{illustr}
e^2 {\cal L}^{2d}_{\rm tree} =  - \frac 1{2} 
\int d^4\theta (\Gamma_1^2 + \Gamma_2^2) =\ -\int  d^4\theta \bar\Sigma \Sigma 
\ \equiv \ \int  d^4\theta \bar\Phi \Phi
 \ee

It is very instructive to derive the effective $2D$ lagrangian {\it directly} elucidating its 
relationship with the SQM effective lagrangian (\ref{Lcomp}) discussed in the previous section.
To do it, consider the original theory  not on $R^2$ and not  on $R^1$, but
rather on $R^1 \times S^1$. Playing with the
length $L$  of the circle, one can interpolate between $1D$ and $2D$ pictures \cite{akhmedov}. 

The Lagrangian (\ref{Lcomp}) was obtained after integrating out the charged fields in $1D$ theory. 
Thinking in $1D$ terms, we have now an infinite number of charged fields representing the coefficients 
in the Fourier series
  \be
  \label{Fourier}
  f(z,t) \ =\ \sum_{n = -\infty}^\infty  f_n(t) e^{i nz/L} \ .
  \ee
 The relevant variables in the effective Lagrangian are still
 zero Fourier modes of the vector potential $\vec{A} \equiv (A_{j = 1,2}, A_3)$ and its superpartners.
The expression (\ref{Lcomp}) is replaced by the infinite 
sum\,\footnote{\,The notation $e_1$ indicates that we are dealing with the coupling constant in $1D$ theory,
$[e_1] \sim m^{3/2}$.}
 \be
 \label{LR1S1}
e_1^2{\cal L}  \ =\
 \left[ 1 +  \sum_{n = -\infty}^\infty \delta h \left(A_j, A_3 + \frac {2\pi n}{ L} \right)\right] (\dot A_j^2
+ \dot A_3^2) 
+\ {\rm other\ terms}
\ee
In the limit $L \to 0$, only one term in the sum survives and we are reproducing the
 previous $1D$ result (with $\delta h = e_1^2/(2|\vec{A}|^3)$). But for large $L \gg e_1^{-2/3}$ all terms
 are essential.
In the limit
 $L \to \infty$, we can actually replace the sum by an integral, $\sum_n \longrightarrow \frac L{2\pi}
\int dA_3$. This integral depends on $A_j$, but not on $A_3$: the expression in square brackets in
Eq.\,(\ref{LR1S1}) gives 
  \be
  \label{htilde}
  \tilde{h} \ =\ 1 + \frac {e_2^2}{2\pi A_j^2}
  \ee
with $e_2^2 = e_1^2L$. This agrees, of course, with (\ref{LbosD2}). 
Actually, in the  limit $L \to \infty $, the effective lagrangian cannot
depend on $A_3$. For large $L$,
 the range where $A_3$ changes is very small, $0 \leq A_3 \leq 2\pi/L$, and the eigenmodes of the
hamiltonian  $\Psi_n(A_3) \sim \exp\{inA_3\}$ with $n \neq 0$ acquire large energy and decouple;
only the mode $n=0$ survives.  To do this
limit carefully, we cannot just set $A_3 = 0$ in Eq.\,(\ref{LR1S1}), however, but 
should perform
the functional integral of $e^{iS}$ [$S$ is obtained from Eq.\,(\ref{Lcomp}) by substituting $\tilde{h}$ 
for $h$]
over  $\prod_t dA_3(t)$ first. Doing this and integrating out also the auxiliary field $D$, we
arrive at the result
   \be
e_2^2{\cal L}_{2D} = \frac 12 g_{jk} \dot A^j  \dot A^k + \frac {i\tilde{h}}{2}
\left({\bar \psi} \dot{ \psi} - \dot{{\bar \psi}}  \psi 
\right) + i \tilde{h} \omega_j^{ab} \dot A^j \bar\psi \sigma^{ab} \psi + \nonumber \\
  \frac{1}{8 \tilde{h}} \left[ (\partial_j \tilde{h})^2 -  \tilde{h}
 (\partial^2 \tilde{h }) \right]
 \left(\bar{\psi}\right)^2 \left(\psi\right)^2\ ,
\label{LKahl}
    \ee
where we have raised the index of the vector
$A^j$ indicating its contravariant
nature, $g_{jk} = \tilde{h} \delta_{jk}$, 
$\sigma^{ab} = \frac i2 \epsilon^{abc} \sigma^c = \frac i2 
\epsilon^{ab} \sigma^3$ ($a,b = 1,2$ ) is the
generator of rotations in the tangent space, and
  \be
\label{omega2}
\omega_i^{ab} = \frac 12 \left[ \delta^a_i \, \partial^b \log{(\tilde{h})} -
\delta^b_i \, \partial^a \log{(\tilde{h})} \right]
   \ee
is the spin connection on a conformally flat
manifold with the natural choice of the  zweibein,
$e_j^a = \sqrt{\tilde{h}} \delta_j^a$.

 When deriving (\ref{LKahl}), we went
 over from the lagrangian ${\cal L}_{1D}$ to 
$2D$ lagrangian density ${\cal L}_{2D} = {\cal L}_{1D}/L$ . (Normally, 
${\cal L}_{1D}$ is a spatial integral of ${\cal L}_{2D}$, but 
we have dealt up to now 
only with the terms depending on zero spatial Fourier modes, in which case the spatial
integral is reduced to multiplication by $L$.)

The lagrangian (\ref{LKahl}) coincides with the standard lagrangian of
K\"ahler supersymmetric sigma model \cite{sigmod} 
in the QM limit. In particular, the coefficient of the 4--fermion term represents 
a $2D$ scalar curvature.\footnote{\,Incidentally, though the bifermion 
term in (\ref{Lcomp}) can be interpreted
in terms of $3D$ spin connection, the 4--fermion term there (before or after  integrating out
 $D$) is {\it not} expressed in terms of $3D$ curvature.}
The full $(1+1)$ effective lagrangian could be obtained
if taking into account the higher Fourier harmonics $\propto \exp\{inz/L\}$
of $A_j(z,t)$ and $\psi_\alpha(z,t)$ in the background.  

 The result (\ref{Kab}) can be readily generalized for an arbitrary non--Abelian 
gauge group. 
The K\"ahler potential depends on $r$ complex chiral superfields $\Phi^A$ and has the same 
sum--over--the--roots structure
as the $1D$ prepotential in Eq.\,(\ref{Fsumj}), 
   \be
\label{Ksumj}
g^2 {\cal K}(\Phi^A) \ =\   \sum_j \left[  \frac 2{c_V} \bar\Phi^{(j)} 
\Phi^{(j)} - 
\frac {3g^2}{4\pi} \ln   \bar\Phi^{(j)}  \ln   \Phi^{(j)}
\right]
\ ,
 \ee 
where $\Phi^{(j)} = \alpha_j(\Phi^A)$.

\subsection{${\cal N}=2:$ Twisted sigma model}

We start again with analyzing Abelian theory. The effective lagrangian involves now 
two complex bosonic variables 
 \be
  \label{fisi}
\sigma = (A_1 + iA_2)/\sqrt{2},\ \ \ \ \ \ \ \
\phi = (A_4 + iA_5)/\sqrt{2} 
\ .
  \ee
One loop calculation brings about a nontrivial metric in the target space 
($\sigma, \bar\sigma, \phi,  \bar\phi $).
This metric can be related
to the SQM 5--dimensional metric by integrating the latter
over $A_3$ by the same token as the K\"ahler metric (\ref{LbosD2})  
was obtained from the metric 
 of the SQM model in the ${\cal N} =1$ case. 
    \be
\label{metrfisi}
e^2 \left. ds^2_{1+1}\right|_{{\cal N} =2} \ =\ 
\left(1 +  
\int_{-\infty}^\infty \frac {dA_3}{2\pi}
\delta h_{0+1} \right)  \ = \nonumber \\
\left[1  + \frac {e^2}{4\pi (\bar\phi \phi + \bar\sigma \sigma) } \right]  
 (2 d\bar\sigma d\sigma
+ 2 d\bar\phi d\phi)\ .  
    \ee
We expect the effective action to have the
$\sigma$ model form. One could worry at this point because
the metric (\ref{metrfisi}) is not
hyper-K\"ahler (the Ricci tensor and the scalar
curvature do not vanish), while hyper-K\"ahler nature of
the metric was shown to be  necessary  for
the standard (1+1) $\sigma$ model to enjoy ${\cal N}= 4$
supersymmetry \cite{Freedman}. In our case, 
${\cal N}= 4$ supersymmetry is there but the metric is not
hyper-K\"ahlerian, and this seems to present a paradox.
The resolution is  that the  $\sigma$ model in hand  
{\it is}  not standard \cite{N2}. 
   Indeed, the bosonic part of the Lagrangian involves besides the standard kinetic term 
$h\left( \partial_\alpha \bar \sigma 
\partial_\alpha \sigma   + \partial_\alpha \bar \phi
\partial_\alpha \phi \right) $  also the "twisted" term 
$\propto \epsilon_{\alpha\beta} \partial_\alpha  \sigma
\partial_\beta \phi $ and $\propto \epsilon_{\alpha\beta}
 \partial_\alpha  \bar\sigma
\partial_\beta \bar\phi$. To understand where the twisted term comes from, consider a 
charged fermion loop in the background
   \be
   \label{backfisi}
   \sigma = \sigma_0 + \sigma_\tau \tau + \sigma_z z,\ \ \ \ 
\phi = \phi_0 + \phi_\tau \tau + \phi_z z
  \ee
  ($\tau$ is the Euclidean time). The contribution to the 
effective action is $\propto \ln \det \| {\mathfrak D}\|$, 
where ${\mathfrak D}$ is the 6--dimensional Euclidean Dirac 
operator, which can be written in the form
   \be
   \label{Dir6} 
   {\mathfrak D}  \ =\ i \frac{\partial}{\partial\tau} + \gamma_3
   \frac{\partial}{\partial z} -i (\gamma_1 A_1 + \gamma_2 A_2 +
   \gamma_4 A_4 + \gamma_5 A_5 )\ .
     \ee     
 Now, if $A_4$ and $A_5$ were absent, we could write 
 ${\mathfrak D} = \gamma_4 (i\tilde{\gamma}_\mu {\cal D}_\mu )$ , with
 $$\mu = 1,2,3,4;\  {\cal D}_4 = \frac{\partial}{\partial\tau} ,\ 
 {\cal D}_3 = \frac{\partial}{\partial z},\ {\cal D}_{1,2}=  - i A_{1,2}; \  
\tilde{\gamma_4} = \gamma_4, \tilde{\gamma}_{1,2,3} = -i\gamma_4 \gamma_{1,2,3} $$
and then use the squaring trick
  \be
  \label{square}
  \det \| {\mathfrak D} \| \ =\ \det \| i \tilde{\gamma}_\mu {\cal D}_\mu \| \ =\ 
\det\,\!^{1/2} \left\| - {\cal D}^2 + \frac i2 
\tilde{\sigma}_{\mu\nu} F_{\mu\nu} \right\|\ ,
   \ee
with $F_{14} = -\partial A_1 /\partial \tau$, etc. The effective action
would be proportional to 
    \be
       \label{intFF}
{\rm Tr} \{\sigma_{\mu\nu} \sigma_{\alpha\beta} \}
F_{\mu\nu} F_{\alpha\beta} \int \frac {d^2p}{4\pi^2} 
\frac 1{(p^2 + 2\bar\sigma \sigma )^2}\ \propto F_{\mu\nu}^2\ ,
  \ee
  which gives the renormalization of the kinetic term while the twisted term does not appear.
 The squaring trick works also in the case where $A_{4,5}$ are nonzero, but do not depend on $\tau,z$. 
Then $2\bar\phi \phi$ is just added to $-{\cal D}^2$ in Eq.\,(\ref{square}) and to $2\bar\sigma \sigma$
 in Eq.\,(\ref{intFF}) leading to Eq.\,(\ref{metrfisi}).
 But in the generic case the fermion determinant cannot 
be reduced to $\det^{1/2} \| - {\cal D}^2 + \frac i2 \sigma_{\mu\nu} F_{\mu\nu} \|$. The basic reason for 
this impasse is that one cannot adequately ``serve" six components of the gradient with only five $\gamma$
 matrices.\footnote{\,By the same reason, the squaring trick does not work for Weyl 2--component fermions in 4 dimensions:
 three Pauli matrices that are available in that case are not enough to do the job.}
  As a result, the extra twisted term in the determinant appears.
  
  We need not perform an explicit calculation here as the twisted and all other terms in the Lagrangian are
 fixed by supersymmetry. The twisted ${\cal N} = 4$ supersymmetric $\sigma$ model was constructed almost 
20 years
 ago \cite{GHR}. At that time it did not attract much attention. Recently, there is some revival of interest 
in the
 GHR model: it happened to pop up in some  string--related problems \cite{brany,DiaSei}. It also pops up as 
the 
effective $(1+1)$ Lagrangian in the case under study.

It was shown that, for ${\cal N} = 4$ supersymmetric generalization to be possible, 
the conformal factor in the metric $h(\bar\sigma, \sigma, \bar\phi, \phi)$ should satisfy the
 harmonicity 
condition
  \be
  \label{harm4}
 \frac {\partial^2 h}{\partial \bar\sigma \partial \sigma } +
 \frac {\partial^2 h}{\partial \bar\phi \partial \phi} \ =\ 0\ .
  \ee
  Obviously, (\ref{metrfisi}) satisfies it everywhere besides the origin. The relationship 
of (\ref{harm4}) to the 5--dimensional harmonicity condition for the metric in the effective 
SQM model (\ref{LO5}) is also obvious. Indeed, integrating a $D$--dimensional harmonic function 
over one of the coordinates like in (\ref{metrfisi}), we always arrive at a $(D-1)$--dimensional
 harmonic 
function.

To construct the full action, consider along with the standard chiral multiplet $\Phi$ satisfying 
the conditions ${\cal D}_\alpha \Phi= 0$ also a {\it twisted} chiral multiplet $\Sigma$ which 
satisfies 
the constraints (\ref{contwist}). As we have seen, the action depending on only $\bar\Sigma$
and $\Sigma$ can be expressed in terms of standard chiral multiplets.
 However, one can write nontrivial Lagrangians involving 
{\it both} $\Phi$ and $\Sigma$. The twisted $\sigma$ model is determined by the expression
  \be
  \label{LGHR}
  {\cal L} \ =\ \int d^2 \theta d^2 \bar\theta 
  \ {\cal K} (\bar\Phi, \Phi; \bar\Sigma, \Sigma)\ ,
     \ee
     where the prepotential ${\cal K}$ satisfies the harmonicity condition,
  \be
  \label{harm4K}
 \frac {\partial^2 {\cal K}}{\partial \bar\Sigma \partial \Sigma } +
 \frac {\partial^2 {\cal K}}{\partial \bar\Phi \partial \Phi} \ =\ 0\ .
  \ee
The condition (\ref{harm4K}) is required if we want the theory to be ${\cal N} = 4$ supersymmetric. 
This is best seen if expressing the lagrangian in components \cite{N2} and observing that the lagrangian
is symmetric under interchange of fermionic variables entering the twisted and untwisted multiplets
only for a harmonic ${\cal K}$. The composition of this discrete symmetry and ${\cal N} =2$ supersymmetry
that is manifest in (\ref{LGHR})  brings about two extra supersymmetries mixing $\phi$ and $\sigma$ with
the fermion components of ``alien'' ${\cal N} =2$ multiplets. 

One of the possible choices for ${\cal K}$ 
(two functions ${\cal K}$ and 
$${\cal K}' \ =\ {\cal K} + f(\bar\sigma, \phi) + \bar f(\sigma, \bar\phi) + g(\sigma, \phi) + 
\bar g(\bar\sigma, \bar\phi)$$
result, up to a total derivative, in one and the same lagrangian.)
leading to the metric (\ref{metrfisi}) is \cite{Hitchin,DiaSei}
   \be
   \label{Kres}
   e^2 {\cal K} \ =\ {\bar\Sigma \Sigma - \bar\Phi \Phi}\  +
   \frac {e^2}{4\pi} \left[ F\left( \frac {\bar\Sigma \Sigma }{\bar\Phi \Phi} \right)  - 
\ln \Phi \ln \bar\Phi \right]\ ,
 \ee
 where
   \be
   \label{Spence}
   F(\eta) \ =\ \int_1^\eta \frac {\ln(1+\xi)}\xi \ d\xi
    \ee
 is the Spence function. This gives besides (\ref{metrfisi}) a twisted term
   \be
 \label{twist} 
 {\cal L}^{\rm twisted} = \  - \frac {e^2}{4\pi(\bar\sigma \sigma + \bar\phi \phi)} 
 \left[ \frac
 {\sigma}{\bar\phi} \epsilon_{\alpha\beta} (\partial_\alpha \bar\sigma)(\partial_\beta \bar\phi) 
 +\frac  {\bar\sigma} \phi 
 \epsilon_{\alpha\beta}(\partial_\alpha \sigma)(\partial_\beta \phi) 
 \right]  
    \ee
in the lagrangian. 
The twisted term is a  2--form $F$. Its external derivative $dF$
can be associated with the torsion. The above--mentioned freedom  
of choice of
${\cal K}$ corresponds to adding to $F$ the external derivative
of the 1-form ``organized'' from the functions 
$f,\bar f, g, \bar g$. The torsion is invariant under such a 
change.

Consider now a generic non--Abelian case. For a simple Lie group of 
rank $r$, the effective Lagrangian is
\be
  \label{Lr}
  {\cal L} \ =\ \int d^2 \theta d^2 \bar\theta 
  \ {\cal K} (\bar\Phi^A, \Phi^A; \bar\Sigma^A, \Sigma^A)\ ,
     \ee
where $A = 1,\ldots,r$ and the expression for ${\cal K}$ is derived exactly in the same
way as in previous cases. We have
  \be
   \label{Kr}
   g^2 {\cal K} \ =\ \sum_j \left\{ \frac 2{c_V} 
\left[\bar\Sigma^{(j)} \Sigma^{(j)} - \bar\Phi^{(j)} \Phi^{(j)} \right] \right.
\nonumber \\ 
\left. -   \frac {g^2}{2\pi} \left[ F \left( \frac {\bar\Sigma^{(j)} \Sigma^{(j)} }
{\bar\Phi^{(j)} \Phi^{(j)}} \right)  - \ln \Phi^{(j)} \ln \bar\Phi^{(j)}
 \right]
\right\}
\ ,
 \ee
where $\Sigma^{(j)} = \alpha_j(\Sigma^A)$, etc. The prepotential (\ref{Kr}) 
satisfies a generalized harmonicity condition
   \be
  \label{harm4Kr}
 \frac {\partial^2 {\cal K}}{\partial \bar\Sigma^A \partial \Sigma^B } +
 \frac {\partial^2 {\cal K}}{\partial \bar\Phi^A \partial \Phi^B} \ =\ 0
  \ee
for all $A,B$.

 \section{\boldmath{$D=3$}\,: K\"ahler and hyper--K\"ahler}

\subsection{${\cal N} =1:$ Dual photon}

The effective lagrangian for $3D$,  ${\cal N} = 2$ (in $3d$ sense) SQED 
depends on only one gauge invariant superconnection  in the reduced 
dimension $\Gamma_3$. Its component expansion (in Wess--Bagger notation) is 
 \be
\label{Gam3}
\Gamma_3 \ = \ A_3 - \frac 12 \epsilon_{\mu\rho\alpha} F_{\mu\rho}\,
 \theta \sigma_\alpha \bar\theta - D \theta \sigma_3 \bar\theta +
\frac 14 (\partial^2 A_3) \theta^2 \bar\theta^2
+ {\rm fermion\ terms}\ ,
 \ee 
where $F_{\mu\rho}$ is $3D$ electromagnetic field ($\mu, \rho = 0,1,2$). The bosonic terms in the
effective lagrangian are
 \be
\label{act3}
{\cal L} \ =\ \int {\cal F} (\Gamma_3) d^4\theta \ =\ h(A_3) \left[ \frac 12 (\partial_\mu A_3)^2 - 
\frac 14   F_{\mu\rho}  F_{\mu\rho} + \frac {D^2}2\right] \ ,
  \ee
where $h = -{\cal F}''/2$.

It is convenient to perform now the {\it duality transformation}. To this end, present the functional
integral corresponding to the lagrangian (\ref{act3}) in the form
 \be
 \label{dualint}
\int \, \prod dF \, d\Psi \, \exp\left\{i \int d^3x \left( {\cal L} + \frac 12 \epsilon_{\mu\rho\alpha} 
 F_{\mu\rho}
\partial_\alpha \Psi \right) \right\}
  \ee
Integrating this over $\prod d\Psi$ brings about the Bianchi constraints $\epsilon_{\mu\rho\alpha}
 \partial_\alpha F_{\mu\rho} = 0$, which are resolved leading to the standard relation
 $ F_{\mu\rho} = \partial_{[\mu} A_{\rho]}$. But let us do  the integral in 
Eq.\,(\ref{dualint})
over  $\prod dF$ first.  We are left with
   \be
  \label{dualZ}
   \prod \, d\Psi  \exp \left\{i \int d^3x \left[ \frac h2 (\partial_\mu A_3)^2 + 
  \frac 1{2h}  (\partial_\mu \Psi)^2 \right] \right\} \ .
   \ee
The integrand in the exponent is the dual lagrangian (the bosonic part thereof). 
The scalar field $\Psi$ is dual photon.

Let us now introduce  the field 
 \be
\label{Bdef}
B \ =\ -\frac {{\cal F}'(A_3)}2
 \ee
 so that $\partial_\mu B = h \, \partial_\mu A_3$. Introducing a
complex variable $\phi = (B + i\Psi)/\sqrt{2}$ we can write ${\cal L}_{\rm dual}$ in the K\"ahler
form $\int d^4\theta \, {\cal K} (\bar\Phi, \Phi)$. The relation of the K\"ahler potential ${\cal K}$  to
the function ${\cal F}$ can be inferred from Eq.\,(\ref{Bdef}). 

For the effective lagrangian of $3D$ SQED, the particular form of the metric and prepotentials 
${\cal F}$, ${\cal K}$ 
can be found in the framework of ``unfolding the ring'' procedure. 
We have to integrate the one--loop correction to the $2D$ metric in (\ref{htilde}) over one
of the components $A_j$ is the same way as we earlier integrated the correction to the $1D$
metric to derive (\ref{htilde}). We obtain
  \be
\label{h3}
e^2 h_{3D} \ =\ 1 + \frac {e^2}{4\pi |A_3|} + \ldots
 \ee
(with 3--dimensional charge $e$). The metric is singular at $A_3 = 0$. This point
 separates here two completely independent sectors in the moduli space (and in the theory !) 
with positive and negative $A_3$.  We will assume for definiteness $A_3$ to be positive. 
The prepotential entering (\ref{act3}) can be restored from
the metric as 
 \be
\label{F3K3}
-e^2{\cal F}(\Gamma_3) \ =\ \Gamma_3^2  + \frac {e^2}{2\pi} \Gamma_3 \ln \Gamma_3 + \ldots \ .
  \ee
The K\"ahler potential depends only on $\Delta =  (\bar\Phi + \Phi)/\sqrt{2}$ and is given at the 
one--loop level
by a similar formula
  \be
  \label{K3}
{\cal K} \ =\ e^2 \left[ \Delta^2 + \frac 1{2\pi} \Delta \ln \Delta \right]\ .
    \ee
The generalization to non--Abelian case is straightforward. The generalized K\"ahler 
potential is
  \be
 \label{K3r}
  {\cal K}(\Delta^A) \ =\ 
g^2 \sum_j \left\{ \frac 2{c_V} 
(\Delta^{(j)})^2   
 -   \frac {3}{2\pi} \Delta^{(j)} \ln \Delta^{(j)}   \right\} + \ldots
\ ,
 \ee
where $\Delta^A =  (\bar\Phi^A + \Phi^A)/\sqrt{2}$ and $\Delta^{(j)} = \alpha_j(\Delta^A)$

This is not  yet the end of the story, however.  Considerations
of supersymmetry alone do not exclude the presence of {\it superpotential}
$\sim {\rm Re} \, \int d^2\theta \, F(\Delta^A) $ on top of the K\"ahler 
potential in the effective lagrangian. Indeed, such a superpotential is generated  
in non-Abelian $3D$ theories  by
{\it nonperturbative} mechanism \cite{3Dinst}. The mechanism is roughly the same
as  the known instanton mechanism for generating superpotential in $4D$ ${\cal N} =1$ SYM
theory with matter \cite{ADS}. In three dimensions, instantons 
are t' Hooft--Polyakov monopoles.
They have two fermion ``legs''  (zero modes) which leads to generation of gluino condensate.
The superpotential can be restored from the condensate. In the simplest $SU(2)$ case 
it has the form 
  \be
 \label{suppot}
  F(\Delta) \ \sim\ g^4 \exp \left\{ -  {2\sqrt{2} \pi \Phi }  \right\} .
  \ee
The superpotential (\ref{suppot}) lifts the degeneracy on the valley. Actually, the
scalar potential 
$U \sim  \exp \left\{ - {2\sqrt{2}\pi (\phi + \bar\phi) }  \right\}$
corresponding to the superpotential (\ref{suppot}) 
(the exponent $2\sqrt{2}\pi (\phi + \bar\phi) = 4\pi A_3/g^2$ is nothing but the 
$3D$ instanton action)
does {\it not} 
have a minimum at a finite value of $\phi$ (as it does not in the massless   ${\cal N} =1$ 
supersymmetric QCD --- this is a typical ``runaway vacuum'' phenomenon). In 
4-dimensional SQCD, this
can be cured by giving a mass to the matter fields: supersymmetric
vacuum would then occur at a finite value of $\phi$. But in the framework of the game we
are playing here, the form of the lagrangian of the descendants 
is dictated by the original $4D$
theory, and we have to conclude that the $3D$ ${\cal N} =1$ sister simply does not exist 
as a consistent theory.

In principle, this all could happen also in the $2D$ ${\cal N} =1$ theory, 
where the appearance
of superpotential is also not excluded by the symmetry considerations. Moreover, 
nonperturbative solutions, the instantons, exist also in the $2D$ case \cite{Wit2D,2Dinst}. 
They appear
in any $2D$ gauge theory involving only adjoint fields due to nontrivial 
$\pi_1$({\small gauge\ group}) by the same token as ordinary BPST instantons 
in four dimensions
appear
due to nontrivial $\pi_3(G)$. Now, $\pi_1[SU(N)] = 0$, but  if only adjoint fields are present,
the gauge group is globally $SU(N)/Z_N$ and involves $N-1$ topologically distinct
noncontractible loops and, correspondingly, $N-1$ different types of instantons.
In the theories involving only adjoint fermions considered in \cite{2Dinst}.

 However, these
instantons do not generate superpotential in this case by two reasons.
\begin{itemize}
\item 
 The minima of classical action are realized on delocalized constant gauge field strength configurations (like
in Schwinger model). The non--Abelian $2D$ instantons do not know about the scalar fields (like monopoles do) 
and cannot lift the degeneracy on the moduli space. 
\item As was shown in \cite{2Dinst}, the instantons involve 
$N-1$ {\it pairs} of fermion zero modes for each  fermion flavor. 
In our case, there are two flavors and an instanton involves  all together $4(N-1)$ instanton legs, which 
 is too much to generate the fermion condensate
and superpotential.  
 \end{itemize}

\subsection{${\cal N}=2:$ Taub--NUT, Atiyah--Hitchin, and their relatives}

The effective lagrangian involves here $4r$ moduli: three components of vector-potential in reduced
dimensions and a dual photon for each unit of the rank. One of the ways to derive the effective
lagrangian there is to determine first the  effective lagrangian for the theory defined on
$R^2 \times S^1$ (it represents a twisted sigma model involving an infinite sequence of the Fourier
modes associated with the circle) and unfold the circle as was explained in details in Sect. 3.
 When the length $L$ of the circle becomes large, we can replace the sum over the modes by the integral.
In the Abelian case, we obtain \cite{seliv}
  \be
 \label{Ld3dodual}
e_3^2{\cal L} \ =\ \left(\frac 12 + \frac {e^2}{8\pi |\vec{A}|} \right)
\left[ (\partial_\mu \vec{A} )^2 + (\partial_\mu \tau )^2 \right] \nonumber \\
- \frac {ie^2}{4\pi} \vecg{\omega}(\vec{A}) \epsilon_{\mu\nu} \partial_\mu \tau \partial_\nu \vec{A}\ ,
 \ee
where $g_3^2 = g_2^2 L$ is the 3--dimensional gauge coupling constant, $\mu = 1,2$
(we have not yet added excited  Fourier modes of the slow variables),   
 and  $\vecg{\omega} (\vec{A}) \equiv   \omega(\vec{A}) = cos \theta \, d\phi$ is a 
1-form which coincides with 
the Abelian connection describing a Dirac monopole in space of $\vec{A}$.
 The variables $\vec{A}$ live on $R^3$ whereas the variable $\tau$ lives on the dual circle,
$0 \leq \tau \leq 2\pi/L$. The second term in Eq.\,(\ref{Ld3dodual}) came from the twisted term  
 (\ref{twist}).

When $L$ is very large, the size of the dual circle is very small
which would normally imply that the excitations related to nonzero Fourier modes of $\tau$
would become heavy and decouple. This is exactly what happened  when we reconstructed 
with this method the $2D$ effective Lagrangian  from
the $1D$ one in Sect. 3.  But in the case under consideration 
it would not be correct just to cross out the terms
involving $\partial_\mu \tau$. The presence of the twisted term 
$\propto \epsilon_{\mu\nu} $ prevents us to
do it.

To understand it, consider a trivial toy model,
 \be
 \label{toy}
 {\cal L} = \frac 12 (\dot{x}^2 + \dot{y}^2) + B x \dot{y} \ \ \Longrightarrow  \ 
\ H = \frac 12 \left[ p_x^2  + (p_y - Bx)^2 \right]\ ,
 \ee
where $x \in R^1$, while $y$ is restricted to lie on a small circle,
$0 \leq y \leq \alpha$. The Lagrangian (\ref{toy}) describes a particle living on a   cylinder and 
moving in a  constant magnetic field . Now, if the magnetic field $B$ were absent, the higher
 Fourier modes of the variable $y$ would be heavy and the low--energy spectrum would be continuous
corresponding to free motion along $x$ direction.
When $B \neq 0$, for {\it each} Fourier mode of the variable $y$, we obtain
 the {\it same} oscillatorial spectrum. Only the position of the center of the orbit and not the energy 
depends on $p_y^{(n)} = 2\pi n/\alpha$.

Thus, we cannot  suppress the variable $y$ in the Lagrangian (\ref{toy}). Likewise,  we cannot  suppress 
the variable $\tau$ in Eq.\,(\ref{Ld3dodual}). What we can do, however, is to trade it to another variable
 using the  duality trick. Performing the same transformations as in the ${\cal N} =1$ case, we arrive
at the lagrangian of the sigma model. 
living on the target space with the metric 
   \be
\label{TAUBNUT}
ds^2 \ =\  \left( 1 + \frac {e^2}{4\pi |\vec{A}|}
\right) d\vec{A}^2 + \frac {
\left(d\Psi - \frac {e^2}{4\pi} \omega  \right)^2}
{\left( 1 + \frac {e^2}{4\pi |\vec{A}|}
\right)} \ .
 \ee
The dual variable $\Psi$ describes the dual photon.

We want to emphasize that the effective lagrangian thus obtained represents a 
{\it conventional} sigma model ---  the twisted term 
disappears after duality transformation. A conventional ${\cal N} =4$ sigma model must be 
hyper--K\"ahler. Indeed,
the metric (\ref{TAUBNUT}) describes a well-known hyper--K\"ahler Taub--NUT manifold
\cite{Eguchi}. 

Consider now SYM theory and let us start with the case of $SU(2)$. The result is
immediately written by substituting  $-2g^2$ for   $e^2$  in all above formulae.  
The metric thus obtained (Taub--NUT with negative mass term) is also hyper--K\"ahler.
However, in contrast to the regular Taub--NUT metric, it is singular at $|\vec{A}|
= 2\pi/g^2$ and represents an {\it orbifold}. Note that the singularity occurs at small
values of $|\vec{A}|$ where the Born--Oppenheimer approximation is not valid anymore.
As was explained earlier, we cannot neglect higher--derivative terms
in the effective lagrangian  in this region. Moreover, the very notion of the effective
lagrangian makes no sense anymore. Still, the presence of singularity in the double--derivative
part of the lagrangian is somewhat irritating.

The remarkable fact is that the singularity actually disappears  when taking into account
{\it nonperturbative} effects associated with instantons (coinciding with 
 't Hooft--Polyakov monopoles). Instantons bring about
the corrections to the metric  
$\sim \exp\{-4\pi n|\vec{A}|/g^2 \}$ ($n$ is the topological charge). 
They are irrelevant in the asymptotics, but
are hundred percent  important  for small values of $|\vec{A}|$: their resummation gives a  {\it smooth}
hyper--K\"ahler Atiyah--Hitchin metric.\,\footnote{
This was the conjecture of Ref. \cite{SW1} confirmed by direct evaluation of the one-instanton contribution
to the metric in Ref. \cite{Tong}. (Multiinstanton contributions need not be calculated. Hyper-K\"ahler nature
of the metric fixes them once the one--instanton contribution is known.)}
There exist an explicit expression for the AH metric. It involves elliptic functions and is not so
simple. An interested reader may look it up in  \cite{AH} and be convinced that its asymptotics
coincide with Taub--NUT, indeed, and that the corrections to this asymptotics are exponential.

Another remarkable fact is that the same AH metric describes the low--energy dynamics of two BPS
monopoles \cite{GM}. In this case, the vector $\vec{A}$ acquires the meaning of the monopole
separation $\vec{r}$ and $\Psi$ of their relative phase. The (smoothened) singularity occurs 
when the distance between the monopoles is of the same order as the size of the monopole cores.
The classical trajectories of the monopoles represent geodesics on the AH manifold.

By the same token as we did it before, we can write the effective lagrangian for an arbitrary
gauge group,
  \be
  \label{L3d}
  g^2{\cal L}\ =\ \frac 12 (\partial_\mu \vec{A}^A )  (\partial_\mu \vec{A}^B ) Q_{AB} + 
\frac 12  J_\mu^A Q^{-1}_{AB} J_\mu^B \ ,
  \ee
  where
  \be
  \label{QiJ}
 Q_{AB} &=& \delta_{AB} - \frac {g^2}{2\pi} \sum_j \frac
 {\alpha_j^A \alpha_j^B}{|\vec{A}^{(j)}|}\ , \nonumber \\
 J_\mu^A &=&  \partial_\mu \Psi^A +
 \frac {g^2}{2\pi} \sum_j \vecg{\omega}(\vec{A}^{(j)})
 \partial_\mu \vec{A}^{(j)} \alpha_j^A \ ,
  \ee
$\vec{A}^{(j)} = \alpha_j(\vec{A}^A) \equiv  \alpha_j^A \vec{A}^A $.

These are asymptotical expressions for the metric. They involve 
singularities and their structure is complicated. However, resummation of instanton corrections
should patch up this singularities. The result of such a resummation gives a smooth hyper--K\"ahler
manifold. Let me give arguments in favor of this conclusion.

{\bf (i)} Consider first the unitary groups \cite{CH}. The Cartan subalgebra of $SU(N)$ consists of 
traceless $N \times N$
diagonal matrices. As far as the effective lagrangian (\ref{L3d}) is concerned, we have four 
such matrices:
diag $(\vec{A}_1, \ldots, \vec{A}_N)$ and diag $(\Psi_1, \ldots, \Psi_N)$, 
$\sum_m \vec{A}_m = \sum_m \Psi_m = 0$.
 There are $N(N-1)/2$
positive roots, $\alpha_{ml}(\vec{A}) = \vec{A}_m - \vec{A}_l,
m < l = 1,\ldots,N$.
Substituting
 it in Eq.\,(\ref{L3d}), we obtain the metric 
   \begin{eqnarray}
  \label{CHmetr}
   ds^2\ = \ A_{ml} d\vec{A}_m d\vec{A}_l + A^{-1}_{ml} \Lambda_m \Lambda_l\ ,
  \end{eqnarray}
where  $A$ is the following $N\times N$ matrix:
  \be
  \label{Aml}
 A_{mm} &=& 1 - \frac {g^2}{4\pi} \sum_{l\neq m} \frac 1{|\vec{A}_m - \vec{A}_l|}\ \ \ \ 
\ \ \ \ \ \ \ \ \ \ ({\rm no\ summation\ over } \ m), \nonumber \\
A_{ml}  &=& \frac {g^2}{4\pi |\vec{A}_m - \vec{A}_l|},\
\ \ \ \ \ \ \ \ \ \ \ \ \ \ \ \ \ \ \ \ \ \ \ \ \ \ (m \neq l)\ ,
 \ee
 and
 $$ \Lambda_m = d\Psi_m +  \frac {g^2}{4\pi} \sum_{l \neq m}
\omega(\vec{A}_m - \vec{A}_l) \ .$$

This metric happens to describe the dynamics of $N$ well--separated BPS monopoles \cite{GM}. 
$\vec{A}_m \equiv \vec{r}_m$ and
$\Psi_i$ are interpreted as positions and phases of individual monopoles. The condition 
$\sum_m \vec{r}_m = 0$
and similarly for phases means that the trivial center of mass motion is separated out. 
The classical dynamics is described by the  following 
equations of motion\,\footnote{\,Here $g$ is interpreted as the monopole magnetic charge. The equations (\ref{eqmotmon})
are  classical as far as the variables $\vec{r}_l$ are concerned, but the quantization of the dynamic 
variables $\Psi_l$ is already carried out. 
The  spectral parameters $q_l$ are quantized to (integer)/$g$ and are interpreted as the 
electric charges of the corresponding dyons.}
 \be
 \label{eqmotmon}
 \ddot{\vec{r}}_l - \frac {g^2}{4\pi} \sum_{m\neq l}^N \frac {\ddot{\vec{r}}_{lm} }{\vec{r}_{ml}} + 
\frac {g^2}{8\pi} \sum_{l \neq m = 1}^N \frac{2\left[\dot{\vec{r}}_{ml} \times \vec{r}_{ml}
\right] \cdot \dot{\vec{r}}_{ml} - \vec{r}_{ml} (\dot{\vec{r}}_{ml}^2) }{r_{ml}^3} \nonumber \\
- \frac {g}{4\pi} \sum_{m \neq l  } (q_{ml}) \dot{\vec{r}}_{ml} \times \frac {\vec{r}_{ml}}{r_{ml}^3} + 
\frac 1{8\pi} \sum_{m \neq l} \frac { q_{ml}^2 \vec{r}_{ml}}{r_{ml}^3} \ = 0\ , \nonumber \\
q_l \ =\ gA^{-1}_{lm} \left[ \dot{\Psi}_m + \frac {g^2}{4\pi}
\sum_{n \neq m} \vecg{\omega}( \vec{r}_{nm} ) \dot{\vec{r}}_{nm} \right] \ =\ {\rm const}\ ,
 \ee
where $\vec{r}_{ml} = \vec{r}_m - \vec{r}_l,\ q_{ml} = q_m - q_l$. The equations of motion for the 
effective lagrangian
(\ref{L3d}) have a similar form, with time derivatives being replaced by $\partial_\mu$ (and $\vec{r}$
by $\vec{A}$). 

The metric (\ref{CHmetr}) is singular
for certain small values of the distances between the monopoles $|\vec{r}_{ml}|$.  
These singularities can be patched, however, and with all probability {\it are} patched by the instanton
corrections.
A following conjecture of existence and uniqueness can now be formulated: 
there is only one smooth
hyper-K\"ahler manifold of dimension $4(N-1)$ ( a {\it generalized} Atiyah--Hitchin manifold) 
 with the asymptotics (\ref{CHmetr}). (I bet there is though, as far as I know, this has not yet 
been proven  
mathematically in an  absolutely rigorous way.)  An explicit expression for the generalized AH 
metric is not known.

{\bf (ii)} 
 $Sp(2r)$. There are $r$ long positive
roots $\alpha_m (\vec{r}) =  \vec{r}_m $
and $r(r-1)$
short positive roots   $\alpha_{ml}(\vec{r}) =
(\vec{r}_m \pm \vec{r}_l)/2$ ($m < l = 1,\ldots, r$ ; $\vec{r}_m$ are mutually orthogonal and
 linearly independent). The
metric reads
\begin{eqnarray}
\label{C}
ds^2  &=& \sum_m (d{\bf r}_m)^2 - \frac{g^2}{4\pi} \sum_{\pm}
\sum_{m<l}
\frac{(d {\bf r}_l \pm d{\bf r}_m)^2}
 {\vert{\bf r}_l \pm {\bf r}_m \vert}
-   \frac{g^2}{2\pi}  \sum_m \frac{(d{\bf r}_m)^2}{r_m} +\ {\rm phase\ part} \nonumber \\
 &\equiv&
Q_{ml} d{\bf r}_m d{\bf r}_l +\ {\rm phase\ part}\ . 
\end{eqnarray}
The full metric is restored from Eqs.\,(\ref{L3d}, \ref{QiJ}).

An important observation is that the
corresponding effective Lagrangian (the QM version thereof)
is obtained from  the effective Lagrangian describing the dynamics
 of $2r+1$ BPS monopoles
numbered by the integers $j=-r,\ldots, r$
by imposing the constraints
\be
\label{A2C}
\vec{r}_{-r} + \vec{r}_{r} \ = \dots = \vec{r}_{-1} + \vec{r}_{1} \ =\ 2\vec{r}_0 \ =\ 0 \ , \nonumber \\
\Psi_{-r} + \Psi_{r} \ = \dots = \Psi_{-1} + \Psi_{1} \ =\ 2\Psi_0 \ =\ 0\ .
 \ee
 We {\it are} allowed to impose these constraints because they are compatible with the equations
 of motion (\ref{eqmotmon}).
The corresponding  metric is hyper-K\"ahler. It has to be, due to ${\cal N} = 4$ supersymmetry, absence
of the twisted term and the theorem \cite{Freedman}. 
One can also prove it more
directly reproducing the result (\ref{C})
by the  hyper-K\"ahler reduction procedure worked out in \cite{GR}. 

{\bf (iii)} $SO(2r+1)$. The system of roots is the same as for $Sp(2r)$ only the long and short roots are 
interchanged: there are now $r(r-1)$ long roots
$(\vec{r}_m \pm \vec{r}_l)/\sqrt{2}$ and $r$ short roots $\vec{r}_m/\sqrt{2}$.
 The metric reads
\begin{eqnarray}
\label{B}
ds^2\ =\
\sum_m (d{\bf r}_m)^2 - \frac{g^2}{2\pi \sqrt{2}} \left[ \sum_{\pm}
\sum_{m<l}
\frac{(d {\bf r}_l \pm d{\bf r}_m)^2}
 {\vert{\bf r}_l \pm {\bf r}_m \vert} + \sum_m \frac{(d{\bf r}_m)^2}{r_m}
\right] \nonumber \\
+\ {\rm phase\ part}\ .
\end{eqnarray}
This metric is obtained from the Gibbons-Manton type  metric for $2r$ BPS monopoles
numbered by the integers $j=-r, \ldots, r,\; j\neq 0$ by
imposing the constraints
  \be
\label{A2B}
\vec{r}_{-r} + \vec{r}_{r} \ = \dots = \vec{r}_{-1} + \vec{r}_{1} \ =\ 0 \ , \nonumber \\
\Psi_{-r} + \Psi_{r} \ = \dots = \Psi_{-1} + \Psi_{1} \  =\ 0\
 \ee
 and rescaling $ds^2$ and $g^2$.
The constraints (\ref{A2B}) are compatible with the equations of motion.

Note that we obtained the effective Lagrangian for $Sp(2r)$ out of that for $SU(2r+1)$ and not 
$SU(2r)$, as one could naively expect in view of the embedding
$Sp(2r) \subset SU(2r)$. Likewise, the moduli space for $SO(2r+1)$ is
obtained out of  $SU(2r)$  and not  $SU(2r+1)$. This is  due to
the fact that
magnetic charges are coupled to coroots rather than roots.

The effective lagrangian for $SO(2r)$ can also be readily written. It can also be interpreted in
monopole terms and the corresponding manifold is related to a generalized AH manifold for the
system of $2r$ monopoles by hyper--K\"ahler reduction accompanied by a certain mass {\it deformation}
(suppressing interactions between certain monopoles) \cite{seliv}.

The constraints (\ref{A2C}), (\ref{A2B}) have the form of ``mirrors'' in the monopole 
configuration space. In the $Sp(2r)$ case, the mirror passes through one of the monopoles while in the
$SO(2r+1)$ case it does not. Such ``mirrors'' appeared earlier in string--related problems and were 
christened
{\it orientifolds} \cite{H}. For me, they just represent graphical pictures describing embedding
of the symplectic and orthogonal groups into unitary ones.

{\bf (iv)} $G_2$. This is the simplest exceptional group. 
 There are three long ( $\vec{r}_1 - \vec{r}_2, \vec{r}_1 - \vec{r}_3, \vec{r}_2 - \vec{r}_3$)
 and three short ($\vec{r}_{1,2,3}$)
positive roots (the constraint $\vec{r}_1 + \vec{r}_2 + \vec{r}_3 =
 0 $  being imposed). The metric reads
 \be
 \label{G}
 ds^2 \ =\ \sum_{m=1}^3 d\vec{r}_m^2 - \frac {g^2}{2\pi}
 \left( \sum_{m>l=1}^3 \frac {(d\vec{r}_m  - d\vec{r}_l)^2}{|\vec{r}_m  - \vec{r}_l|}  + 3 \sum_{m=1}^3 
\frac {d\vec{r}_m^2}{|\vec{r}_m |} \right) + \ldots\ .
  \ee
It can be obtained out of the metric for $Sp(6)$
  \be
 \label{Sp6}
 ds^2 \ =\ \sum_{m=1}^3 d\vec{r}_m^2 -
 \frac {g^2 }{4\pi}
 \left( \sum_\pm \sum_{m>l=1}^3 \frac {(d\vec{r}_m  \pm d\vec{r}_l)^2}{|\vec{r}_m  \pm \vec{r}_l |}  +
 2 \sum_{m=1}^3 \frac {d\vec{r}_m^2}{|\vec{r}_m |} \right) + \ldots
  \ee
  by rescaling and  imposing the [compatible with $Sp(6)$ equations of motion] constraints
\be
  \label{C3-2-G2}
\vec{r}_1 + \vec{r}_2 + \vec{r}_3 \ =\ 0\ , \ \ \ \ \ \ \
\Psi_1 + \Psi_2 + \Psi_3 \ =\ 0\ .
  \ee
This ``3--fold mirror'' is (can be called) an orientifold of new type.
  Again, we obtained ${\cal L}_{\rm eff}$ for  $G_2$ out of   ${\cal L}_{\rm eff}$ for  $Sp(6)$, 
though    
$G_2$ is embedded not into
$Sp(6)$, but into the dual algebra $SO(7)$.

The effective Lagrangian for $F_4$  can be related to the moduli space of
26 monopoles. (26 is the lowest dimension of a unitary group where
$F_4$ can be embedded. This follows from the fact that the representation 
{\bf 26} of $F_4$ has the lowest dimension.) $E_6$ can be embedded into
$SU(27)$ and hence the corresponding effective Lagrangian is
 related to the moduli space of
27 monopoles. Now, the shortest representation in $E_7$ has the
dimension 56 and we need at least 56 monopoles in this
case. Finally, $E_8 \subset SU(248)$ and we need 248 monopoles. The moduli
space of 248 monopoles can also be used as a universal starting point
to describe the dynamics
of $F_4, E_6$ and $E_7$, if
following the chain of embeddings $F_4 \subset E_6 \subset E_7 \subset E_8
\subset SU(248)$.

The  explicit formulae we have written refer to the asymptotic region
 where nonperturbative effects are suppressed. The corresponding metrics
involve singularities at small $|\vec{r}^{(j)}|$.
Like for the $SU(N)$ case, a very reasonable conjecture is that  these
singularities are sewn up by instantons  for any simple Lie group giving a unique
 smooth hyper-K\"ahler metric with the asymptotics
    \be
   ds^2 \ =\ d\vec{r}^A Q_{AB} d\vec{r}^B + \ldots
   \ee
It is  natural to conjecture that this metric is obtained
from the multimonopole Atiyah-Hitchin
metrics by the same hyper-K\"ahler reduction procedure as above. To the best of my knowledge,
 hyper--K\"ahler manifolds thus obtained were
not studied before by mathematicians.  

\section{Non--renormalization theorems}
 There are several proofs of the well--known fact that two-- and higher--order corrections to
the $\beta$ function in the $4D$ ${\cal N} = 2$ supersymmetric Yang--Mills theory vanish.
 We will discuss here two such proofs: {\it (i)} diagrammatic (historically, this was the first) and 
{\it (ii)} the one following from holomorphy.

{\it Supergraphs. } 
The diagrammatic proof is based on the technique of supergraphs. The simplest nonrenormalization 
theorem states that all loop corrections to the {\it superpotential} (the term $\int \, d^2\theta F(\Phi_i)$ 
in the lagrangian, $\Phi_i$ are chiral superfields) vanish. We refer the reader to the textbooks
\cite{WB,Siegel,West} for its proof,  recall
in some more details how it is done for gauge couplings (following Refs.\cite{ShiVa}), the subject of our
interest here. 

Consider for simplicity SQED.\footnote{\,Generalization to 
non--Abelian case is relatively straightforward, but it 
involves some subtleties associated with infrared singularities of the theory \cite{ShiVa,SmiVa}
 which we do not like to discuss here.}
We explained above [see Eq.\,(\ref{detratio})] how the one--loop correction 
to the effective action in any dimension can be evaluated. It does not vanish here. 
The effective charge in ${\cal N} =1$ theory is given by
 \be
 \label{eren4}
  \frac 1 {e^2_{\rm phys}} \ =\ \frac 1{e_0^2} + \frac 1{4\pi^2} \ln \frac {\Lambda_{UV}} {m_0} + \ldots\ ,
  \ee
where $m_0$ is the {\it bare} charged fields mass.
\begin{figure}
   \begin{center}
        \epsfxsize=170pt
        \parbox{\epsfxsize}{\epsffile{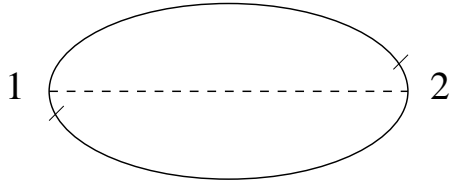}}
        \vspace{5mm}
    \end{center}
\caption{Two-loop contribution to the effective action. 
Solid lines are chiral field superpropagators $\langle \Phi \bar \Phi \rangle$ evaluated in the classical
gauge background and the dashed
line is the propagator of the (quantum part of the) vector superfield $V$. The bar on the solid line marks
the $\bar \Phi$ end.} 
\label{2loop}
\end{figure}

The relevant 
two--loop graph is drawn in Fig. \ref{2loop} (there are actually two such supergraphs giving the same
contribution with the superfields $S$ or $T$ in the loop).
 According to the supergraph Feynman rules \cite{Siegel,West},  each vertex
involves the integral $\int d^8z = \int d^4x  \, d^2\theta d^2\bar\theta $ and the whole contribution
of the graph in Fig. \ref{2loop} is
 $ \int d^4x_1 \, d^2\theta_1  d^2\bar\theta_1  K$, where
$$ {\cal K} = \frac i2 \int d^8z_2 \langle \Phi_1 \bar \Phi_2 \rangle  \langle  \Phi_2 \bar \Phi_1  \rangle
\langle v_1 v_2 \rangle \ .$$   
 Here $\Phi$ stands for charged chiral superfields $S,T$, $v$ is quantum vector superfield
 and $\langle \Phi_1 \bar \Phi_2 \rangle, \ \langle v_1 v_2 \rangle$  are quantum superpropagators 
 evaluated in external background $V_{\rm cl}$.  Now, $ \langle v_1 v_2 \rangle$ does not depend
on external field and on its gauge. The charged field propagators are gauge--dependent:
    \be
\langle S_1 \bar S_2 \rangle \to e^{i\Lambda_1} \langle S_1 \bar S_2 
\rangle e^{-i\bar\Lambda_2}\\ \nonumber
\langle T_1 \bar T_2 \rangle \to e^{-i\Lambda_1} \langle T_1 \bar T_2 \rangle e^{i\bar\Lambda_2}\ .
  \ee
The point is, however, that the {\it integrand} $K$ is gauge--independent and should thereby be
{\it locally}\,\footnote{\,Locality follows from the presence of 
an infrared cutoff (nonzero mass) in the theory.}
expressed via the gauge--invariant superfield $W_\alpha$. 
But $W_\alpha$ is a chiral superfield and
the integral over $d^2\bar\theta$ of any function of $W$ vanishes.  Therefore 
$\int d^2\theta d^2\bar\theta \, {\cal K} =0$
Q.E.D. The same reasoning apply also to an arbitrary multiloop 
graph.\footnote{\,To be quite precise, the integrand could depend 
on {\it both} $W$ and $\bar W$,
in which case the integral  $\int d^2\theta d^2\bar\theta \, {\cal K}$ need not vanish. One can be
convinced, however, that such contribution is a supersymmetric generalization of higher derivative
$\sim F^4$  terms in the Euler--Heisenberg effective lagrangian. {\it Such} corrections to the
effective lagrangian are present, indeed, but this does not affect the renormalization of the gauge
coupling.}

We hasten to comment that this does {\it not} mean that multiloop contributions to $\beta$ 
function in ${\cal N} =1$ supersymmetric QED vanish. Higher loops appear when expressing
$m_0$ entering Eq.\,(\ref{eren4}) into $m_{\rm phys}$. As was already mentioned above, physical
mass {\it is} renormalized in spite of the fact that the mass term in the Lagrangian is not. 
Indeed,
the physical mass is defined as the pole of the fermion propagator $\propto 1/(Z/\!\!\!p - m_0)$, 
where $Z$
describes the renormalization of the {\it kinetic} term 
$$\propto \int d^2\theta d^2\bar\theta \left( \bar S e^V S + \bar T e^{-V} T \right) \ . $$
We have $m_0 = Z  m_{\rm phys}$ what leads to an exact relation expressing the charge
renormalization via the matter $Z$ factor,
 \be
\frac 1 {e^2_{\rm phys}} \ =\ \frac 1{e^2_0} + \frac 1{4\pi^2} \ln 
\frac {\Lambda }{m_{\rm phys} } - \frac  1{4\pi^2} \ln Z \ . 
  \ee
In particular, using knowledge of $Z$ at the one--loop level 
 \be
  \label{mrenorm}
 Z\ = \ 1 -  \frac {e_0^2}{4\pi^2}  \ln \frac \Lambda {m_{\rm phys}} 
  \ee
we obtain the two--loop renormalization of the charge
  \be  
\label{ephys}
\frac 1{e_{\rm phys}^2} \ =\  \frac 1{e_0^2} + 
\frac 1{4\pi^2} \ln \frac {\Lambda}{m_{\rm phys}} 
+ \frac {e_0^2}{16\pi^4} \ln \frac {\Lambda}{m_{\rm phys}} + \ldots 
   \ee

Now, ${\cal N} = 2$ supersymmetric electrodynamics involves an extra neutral chiral superfield
$\Upsilon$. The Lagrangian involves its kinetic term and the extra superpotential term
$\propto \int d^2\theta \, \Upsilon ST $.  The latter is not renormalized: this is the standard
$F$ term nonrenormalization theorem. The point is that this superpotential term is related
by extended supersymmetry to the charged field kinetic term. Hence, nonrenormalization of the
superpotential {\it implies} in  ${\cal N} = 2$ theory nonrenormalization of the kinetic term,
which implies the absence of the mass renormalization. In other words, in  ${\cal N} = 2$ theory,
$m_{\rm phys} = m_0$ and hence only the first term in the $\beta$ function survives. All 
 ${\cal N} = 2$ theories with vanishing 1--loop contribution to the  $\beta$ function are 
finite.  The ${\cal N} = 4$ SYM theory belongs to this class.

The proof just given uses the formalism of ${\cal N} =1$ supergraphs.  ${\cal N} = 2$ symmetry is used
indirectly via requirement of equality of renormalization factors for standard the kinetic and the superpotential
terms. One can also define and calculate supergraphs in  ${\cal N} = 2$ (harmonic) superspace. In this case,
the absence of the corrections is manifest \cite{kniga}.     

\vspace{.5mm} 

{\it Holomorphy}. An alternative elegant proof comes from the analysis of the Seiberg--Witten 
effective lagrangian.
As was mentioned in the beginning, ${\cal N} =2$ supersymmetry dictates the form (\ref{LSW}) 
for the effective
lagrangian where $F({\cal W})$ is a {\it holomorphic} function of the  ${\cal N} =2$ superfield
(\ref{Wcal}) (the Abelian version thereof). 
Consider this function for large ${\cal W}$. Going around the large circle
( multiplying ${\cal W}$ by $e^{2i\pi}$ ), we should obtain the 
{\it same} theory.\footnote{\,Actually, it is sufficient to 
multiply  ${\cal W}$ by $e^{i\pi}$. It is ${\cal W}^2$
rather than    ${\cal W}$ which has direct physical meaning, 
the lowest component of ${\cal W}^2$
coinciding with the true moduli $u = {\rm Tr}\, \phi^2$. }
Knowing that in the asymptotics 
   \be
  {\cal L}_{\rm eff} \sim  {\rm Re}  \,    
\int d^4\theta  \,  \frac 1 {2g^2 ({\cal W})} \, {\cal W}^2 
  \ee
(That was written in non--Abelian $SU(2)$ theory. The same expression with 
$e^2({\cal W})$ substituted
for $g^2({\cal W})$ holds  in ${\cal N} =2$ SQED),  
only two possibilities are allowed:  {\it (i)} $g^2 ({\cal W})$ is a constant (this possibility is
realized for ${\cal N} =4$ non--Abelian gauge theories) or {\it (ii)} . $g^{-2} ({\cal W})$ involves
a term $\sim \ln {\cal W}$, which corresponds to one--loop renormalization. When multiplying
${\cal W}$ by $e^{2i\pi}$, the logarithm is shifted by an imaginary constant. This gives a change
$\sim {\rm Im}\, \int d^4 \theta \, {\cal W}^2$ in the lagrangian, which is a { total derivative}
($\theta$ term). In Abelian theory, $\theta$ term is never relevant. In non--Abelian theory it {\it might}
have been relevant, but it does not in this case: one can be convinced that multiplying
 ${\cal W}^2$ by $e^{2i\pi}$ amounts to the shift $\theta \to \theta + 4\pi$. 

 Higher--order coefficients $\beta_2 , \beta_3$, etc should vanish. Were e.g. $\beta_2$ nonzero, the 
coefficient $g^{-2} ({\cal W})$ in  ${\cal L}_{\rm eff}$ would involve the contribution
$\sim \ln \left|1 + c  \ln (\Lambda_{\rm UV}/ {\cal W} ) \right| $, which is not holomorphic and 
 not allowed. On the other hand,
nothing prevents  the function $f({\cal W})$ from  having 
contributions $\sim {\cal W}^{-n}$, which vanish  in the asymptotics. 
 Indeed,   $f({\cal W})$ {\it does} involve such contributions brought about by  instantons \cite{SW}.

\vspace{1mm}

 We have seen that low-dimensional sisters of $4D$ ${\cal N} =2$ have similar properties: the perturbative
corrections to the effective lagrangia vanish beyond one loop. In sect. 2.2 we explained why: extended
supersymmetry dictates a special form of prepotentials. In $1D$, resp. $2D$ theories, the
prepotentials in (\ref{act5}), resp. (\ref{LGHR}) living in $R\!\!\!\!R^5$, 
resp.  $R\!\!\!\!R^4$  must be harmonic functions of their arguments .
For  $3D$ theories, extended supersymmetry requires the metric to be hyper--K\"ahler. In the asymptotics,
the metric (\ref{TAUBNUT}) involves, indeed a   harmonic function $1 + e^2/(4\pi |\vec{A}|)$ living
 in $R\!\!\!\!R^3$.
A more detailed analysis shows that the harmonicity follows from the hyper--K\"ahler nature of the metric
(i.e. extended supersymmetry) and from its $U(1)$ isometry corresponding to shifting the phase 
$\Psi$ (this isometry shows up  in the 
asymptotics). Actually, in the cases when such an isometry is present, the K\"ahler potential 
of a hyper--K\"ahler metric can be obtained from a certain 3-dim harmonic prepotential by Legendre
(physically - by duality) transformation \cite{Hitchin,seliv}.
\footnote{\,The full Atiyah--Hitchin metric, which involves besides one loop also nonperturbative
instanton corrections does not
have this isometry and cannot be expressed via a 3-dim harmonic function. It can be expressed, however,
via certain more complicated generalized harmonic functions \cite{AHharm}. 
Their physical meaning is yet to
be revealed.}     

Now, in $4D$ theories the moduli space is   $R\!\!\!\!R^2 \equiv C\!\!\!\!C^1$ 
and harmonicity there is the same as 
analyticity\,! In other words, the  proof of the $4D$ nonrenormalization theorem based on 
holomorphy has  direct low--dimensional counterparts. Nonrenormalizability is a family property
of all sisters.

What about the diagrammatic proof ?

The $4D$ diagrammatic proof quoted above involved two parts: {\it (i)} the ${\cal N} =1$
nonrenormalization theorem and {\it (ii)} the 
${\cal N} = 2$ relationship between the kinetic and superpotential
terms. This relationship holds also in low dimensions, but there is no 
${\cal N} =1$ nonrenormalization theorem anymore. Indeed, the theorem was based on the the fact that
the $4D$ effective lagrangian had the form $\int d^2\theta \, W^2$, while the two and higher loop supergraphs
suggested the form $\int d^2\theta  d^2\bar\theta \, X(W, \bar W)$, which could be only reconciled if $X=0$. But
in low dimensions, the effective lagrangian does not have a chiral form, but represents an integral 
 $\int d^2\theta  d^2\bar\theta$ of a local density depending not on $W$, but rather on
superconnections $\Gamma_k$ in reduced dimensions [see e.g. Eq.\,(\ref{act1})]. This can well be reconciled
with what follows from the diagram in Fig. 2. 

Indeed, direct component calculations of the two--loop corrections to the effective action in the 
$D=1, \ {\cal N} =1$ Abelian theory showed that they do not vanish \cite{Abel}. 
One obtains instead of Eq.\,(\ref{hab})
  \be
\label{hab2}  
e^2h(\vec{C})  \ =\  1 + \frac {e^2} {2|\vec{C}|^3}  - \frac {3e^4}{4|\vec{C}|^6 } + \ldots 
 \ee  
 for the metric. In addition, the two--loop contribution is not related to any $Z$-factor, like it is the
case in four dimensions:  the latter just cannot be defined in quantum mechanics.

In Ref.\cite{Abel}, the result (\ref{hab2}) was obtained after rather cumbersome calculations 
where the contribution of several graphs was added. Using ${\cal N} = 1$ supergraph technique, 
only one graph in Fig. 2 should be evaluated and the calculation is rather simple \cite{SmiVa}. 
In ${\cal N} = 2$ theory, a similar graph with the exchange of $\Upsilon$ field should be added.
It has exactly the same structure and gives exactly the same contribution, but with the opposite
sign. The cancellation is manifest. Unfortunately, this simple cancellation pattern does not hold
at the 3--loop level and higher. Again, one obtains zero after adding up several different
supergraphs. In other words, it is hardly possible to prove nonrenormalization of ${\cal N} = 2$
theories using the formalism of ${\cal N} = 1$ supergraphs. 

On the other hand, it is very reasonable to suppose that the diagrammatic proof of
non--renormalizability based on the technique of harmonic supergraphs \cite{kniga} can be extended
to low dimensions. This question is currently under study.

\section{Conclusions}
To make distinction with the Introduction where main results concerning the nature and character
of different sisters were outlined in words and the main body of the paper where the relevant formulae 
were written, we give here the same information in the 
 form of a table.

\begin{table}[ph]
\caption{Pure SYM: the family of effective theories.\vspace*{3pt} }
{
\begin{tabular}{|c|p{5.5cm}|p{5cm}|}
\hline
\ & \ & \ \\[-2mm]
\  &\qquad \qquad \qquad ${\cal N} =1$ &  \qquad \qquad\qquad  ${\cal N} =2$ \\[1mm]
\hline 
\ & \ & \ \\[-2mm]
$D =1$ & Symplectic $\sigma$ model of the first kind & Symplectic $\sigma$ model of the second kind \\[1mm]
\hline
\ & \ & \ \\[-2mm]
$D =2$ & K\"ahler $\sigma$ model & Twisted $\sigma$ model \ (GHR)\\[1mm]
\hline
\ & \ & \ \\[-2mm]
$D =3$ & K\"ahler $\sigma$ model with superpotential.  Run-away vacuum & Hyper--K\"ahler $\sigma$ model \\[1mm]
\hline
\ & \ & \ \\[-2mm]
$D =4$ & No moduli space. Discrete vacua & SW effective theory \\[1mm]
\hline
 \end{tabular}}

\end{table}

\section*{Acknowledgments}

I thank  E. Akhmedov, E. Ivanov, and A. Vainshtein for collaboration
and many  illuminating discussions. Special thanks are due to A. Vainshtein who read the
draft and made many very useful comments. 

{\it  My thanks are also due to my friend and collaborator Konstantin Selivanov. Alas, 
 Kostya 
is not with us anymore. He died young, as Ian did. This paper is a tribute to the memory
of Ian and to the memory of Kostya.}

\end{document}